\documentclass[aps,pra,reprint,twocolumn,superscriptaddress,longbibliography]{revtex4-2}

\usepackage{graphicx}
\usepackage{xcolor}
\usepackage{dcolumn}
\usepackage{bm}
\usepackage{amsfonts}
\usepackage{amsmath}
\usepackage{amssymb}
\usepackage{amsthm}
\usepackage{natbib}
\usepackage{physics}
\usepackage{comment}
\usepackage{appendix}
\usepackage[normalem]{ulem}
\usepackage{hyperref}
\usepackage{orcidlink}
\hypersetup{
  colorlinks=true,
  linkcolor=blue,
  filecolor=blue,
  citecolor=blue,   
  urlcolor=cyan,
}

\usepackage[capitalise]{cleveref}

\begin{document}
\title{Long-Range Interacting Many-Body Systems in the Irrep Basis}
\author{Ivy Pannier-Günther\orcidlink{0009-0003-9970-1658}}
\email[]{igunther@unm.edu}
\affiliation{Center for Quantum Information and Control, Department of Physics and Astronomy, University of New Mexico, Albuquerque, New Mexico 87131, USA}
\author{Andrew Kolmer Forbes\orcidlink{0009-0003-8730-8007}}
\email[]{aforbes@unm.edu}
\affiliation{Center for Quantum Information and Control, Department of Physics and Astronomy, University of New Mexico, Albuquerque, New Mexico 87131, USA}
\author{Pablo M. Poggi\orcidlink{0000-0002-9035-3090}}
\affiliation{Department of Physics, SUPA and University of Strathclyde, Glasgow G4 0NG, United Kingdom }
\affiliation{Center for Quantum Information and Control, Department of Physics and Astronomy, University of New Mexico, Albuquerque, New Mexico 87131, USA}
\author{Ivan H. Deutsch\orcidlink{0000-0002-1733-5750}}
\affiliation{Center for Quantum Information and Control, Department of Physics and Astronomy, University of New Mexico, Albuquerque, New Mexico 87131, USA}
\date{2025-05-13}

\begin{abstract}
Spin models featuring infinite-range, homogeneous all-to-all interactions can be efficiently described due to the existence of a symmetry-restricted Hilbert subspace and an underlying classical phase space structure. However, when the permutation invariance of the system is weakly broken, such as by long- but finite-range interactions, these tools become mathematically invalid. Here we propose to approximately describe these scenarios by considering additional many-body subspaces according to the hierarchy of their coupling to the symmetric subspace, defined by leveraging the structure of irreducible representations (irreps) of the group $SU(2)$. We put forward a procedure, dubbed ``irrep distillation," which defines these additional subspaces to minimize their dimension at each order of approximation. We discuss the validity of our method in connection with the occurrence of quantum many-body scars, benchmark its utility by analyzing the dynamical and equilibrium phase transitions, outline its phenomenology, and compare its use-cases against other approximations of long-range many-body systems.
\end{abstract}

\maketitle
\section{Introduction}
The complexity of quantum many-body systems is of interest from both natural and computation perspectives. 
The mathematical description of such complex systems depends strongly on the range of the interactions, from infinite-range all-to-all couplings as in the Lipkin-Meshkov-Glick model~\cite{Lipkin65} and Tavis-Cummings Hamiltonian~\cite{Tavis68}, to nearest-neighbor interactions such as those in the PXP~\cite{Turner18,Ivanov24} and standard Ising models~\cite{Ising25}. In each of these two extremes, there are geometries for which analytic simplifications permit exact solutions, even at large system-sizes. But for long- but not infinite-range multi-body interactions, the lack of symmetry typically eludes integrability and confounds most exact solving techniques. The goal of this work is to develop new tools to describe these long-range quantum many-body systems and broaden our understanding of the resulting phenomenology.

The range-dependent behavior of quantum many-body physics is especially evident in spin systems as exemplified by the 1D transverse field Ising model which is integrable for infinite-range and zero-range (nearest neighbor) models, but not integrable elsewhere. Permutation-symmetric spin Hamiltonians, in the limit of infinite-range interactions, reduce to collective-spin models representing a single-body degree of freedom and a block-diagonal Hamiltonian that strongly constrains the dynamics. However, any violation of permutation symmetry explodes the dynamically accessible Hilbert space and renders the system irreducibly many-body in description. In many cases the finite-range of interactions is accompanied by a collapse of integrable dynamics into non-integrability and even chaos; the transverse field Ising model is known to be chaotic in this regime~\cite{Russomanno21,Kim14}. While exact integrability can be lost, perturbation theory informs us that sufficiently minor deviations from infinite-range interactions at finite system sizes should still approximate their collective counterparts' dynamics, especially in terms of collective observables like net polarization, spin-number, and entanglement quantifiers. Likewise, reduced-state evolution on finite subsystems in long-range systems are independent of local interactions\cite{Mattes25}.

The near integrable behavior of long-range (but not infinite-range) Ising models under specific initializations was elucidated in recent theoretical developments by Lerose {\em et al.}~\cite{Lerose23} who explained the retention of collective dynamical order by characterizing the subset of eigenstates responsible for high co\"operativity. The strong/weak eigenstate thermalization hypothesis (ETH) predicts that all (strong) or nearly all (weak) eigenstates of chaotic Hamiltonians should be indistinguishable from thermal states according to local measurements \cite{Deutsch18}, and the vanishing fraction of eigenstates which violate the strong ETH are called quantum many-body scars (QMBS) \cite{Serbyn21}. Lerose {\em et al.} demonstrated generic QMBS across a wide class of long-range time-independent spin-Hamiltonians; these QMBS are characterized by low entanglement-entropy, high spin-number, and a high spectral gap between each scar state and neighboring eigenstates outside the QMBS. In the limit of infinite-range interactions, the QMBS reduces to the permutation-symmetric eigenstates of the corresponding collective system.

In this work, we consider a novel scheme of approximation for long-range interacting spin-Hamiltonians which generates a perturbation-theoretic QMBS.  By extending the corresponding subspace from the collective system using $\mathcal{O}(N)$ matrix elements in an efficient basis, we exploit an alternative representation of the Hilbert space, not as a tensor product of $N$ spin-1/2 particles, but as a {\em direct sum of degenerate irreducible representations of $SU(2)$}. We analytically ``distill" these irreducible representations to truncate the Hilbert space down to the nondegenerate minimal subspace necessary to obtain the long-range system as a first-order perturbation on the infinite-range system. 

This article is organized as follows. In Sec. \ref{sec:form} we establish the foundational formalism, including a description of the system of interest, a transverse Ising model with tunable interaction strength and range, which generically exhibits chaos and QMBS. Here we define the decomposition of Hilbert space into irreducible representations of $SU(2)$. By constructing the Hamiltonian in this basis we observe patterns in its block structure, which motivates Hilbert space truncation using an efficient approximation scheme which we dub irrep distillation. In Sec. \ref{sec:app} we apply this formalism in various applications, including a detailed discussion of QMBS, dynamical quantum phase transitions, and the study of the dynamics of nonclassical states, going beyond what is possible with semiclassical approaches. In Sec. \ref{sec:disc} we further explicate the special features of our formalism, defining its niche among the canonical approximations of long-range many-body systems. We explain the unique tensor product structure afforded by irrep distillation, which allows us to describe long-range many-body systems as consisting of two degrees of freedom, thereby generalizing the single-body collective spin description of infinite-range systems. We summarize and provide an outlook for future studies in Sec. \ref{sec:C&O}.

\section{Formalism}\label{sec:form}
\subsection{Model}\label{sec:model}
We consider a class of Ising Hamiltonians with power-law decaying longitudinal interactions and a symmetric transverse field parameterized as
\begin{equation}
 \hat{H}(s,\alpha)=\frac{s-1}{2}\sum_{j=1}^N \hat{\sigma}_x^{(j)}-\frac{s}{4\mathcal{N}}\sum_{j,k=1}^N\frac{\hat{\sigma}_z^{(j)}\hat{\sigma}_z^{(k)}}{|j-k|^\alpha}.\label{eq:Ham1}
\end{equation}
We denote the relative strength of the interactions over the field by a single parameter $s\in[0,1]$, to ensure constant energy scale $||\hat{H}||$ for all parameter values. 
The parameter $\alpha\in[0,\infty)$ describes the power-law decay, or the range of the interactions, and we further guarantee energy scale invariant to $\alpha$ by normalizing with $\mathcal{N}\equiv \frac{1}{N-1}\sum_{j,k}|j-k|^{-\alpha}$, the Kac normalization~\cite{Kac69}. Thus, we have extensive energy scaling with system size $N$, and otherwise totally invariant to parameters. This model arises naturally in ion traps~\cite{porras2004} and Rydberg atom arrays~\cite{Labuhn16} with Van der Waals and dipole-dipole interactions, and many-body dynamics for such long-range interacting systems have been studied in a variety of applications~\cite{schachenmayer2013,richerme2014non, Zhang17,Pappalardi18,Zunkovic18,defenu2024}.

In the limit $\alpha\to 0$, \cref{eq:Ham1} reduces to a Lipkin-Meshkov-Glick (LMG) model~\cite{Lipkin65},
\begin{align}
 \hat{H}(s,0)&=\frac{s-1}{2}\sum_{j} \hat{\sigma}_x^{(j)}-\frac{s}{4N}\sum_{j,k}\hat{\sigma}_z^{(j)}\hat{\sigma}_z^{(k)}\nonumber\\&=(s-1)\hat{J}_x-\frac{s}{N}\hat{J}_z^{2},\label{eq:LMG}
\end{align}
where 
\begin{equation}
 \hat{J}_\gamma = \frac{1}{2}\sum_{i=1}^N \hat{\sigma}_\gamma^{(i)}
\end{equation}
are the components of the collective spin angular momentum satisfying the usual $SU(2)$ commutation relations. In the case of \cref{eq:LMG}, the system commutes with total angular momentum $[\hat{H}(s,0),\hat{J}^2]=0$, and so the LMG eigensystem has the following quantum numbers,
\begin{align}
 \hat{H}(s,0)\ket{n}&=E_n\ket{n},\label{eq:LMGs1}\\
 \hat{J}^2\ket{n}&=J_n(J_n+1)\ket{n}.\label{eq:LMGs2}
\end{align}
That the energy eigenstates are jointly eigenstates of a global operator brings analytical benefits through the expression of the Hamiltonian in the corresponding basis of collective states. This natural basis, in fact, is the Dicke basis, which decomposes into irreducible representations of $SU(2)$, which we will discuss in the next section.

\begin{figure}
\centering
\includegraphics[width=\linewidth]{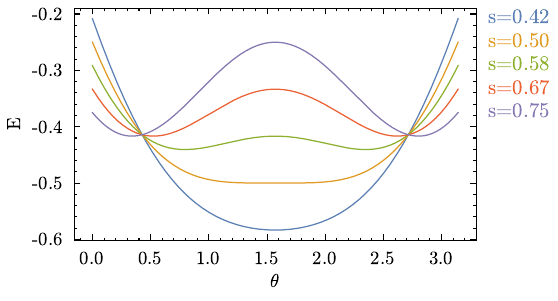}
\includegraphics[width=0.32\linewidth]{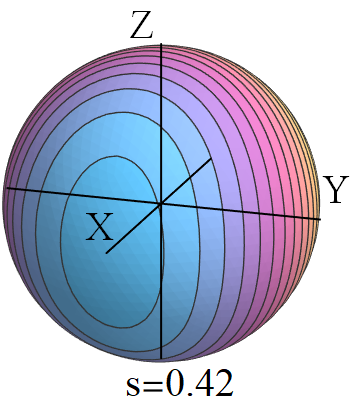}
\includegraphics[width=0.32\linewidth]{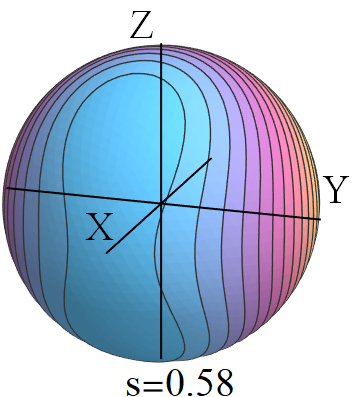}
\includegraphics[width=0.32\linewidth]{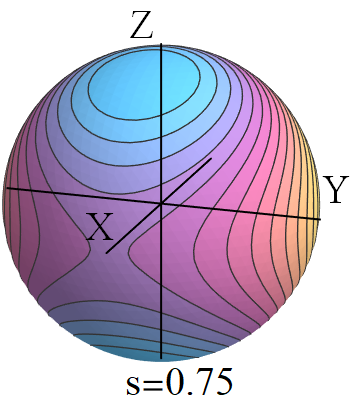}
\caption{\label{fig:classicalPS}The energy function $E(s;\theta,\phi)$ of the collective spin LMG, from \cref{eq:classE}, depicted in cross-section at $\phi=0$, which governs the mean-field dynamics. The bifurcation of the single well to a double well gives rise to a ground-state quantum phase transition (GQPT) at the critical points $s_{GQPT}=\frac{1}{2}$ and a dynamical quantum phase transition (DQPT) as $s_{DQPT}=\frac{2}{3}$. Below, the energy function on the sphere, showing the split of the single-well potential into two wells an corresponding mean-field paramagnetic/ferromagnetic phases.}
\end{figure}

The transverse field Ising model exhibits quantum phase transitions (QPT) between ferromagnetic and paramagnetic phases defined by order parameters in the ground state~\cite{Munoz23} and through dynamical order parameters associated with time-averaged correlation functions~\cite{Heyl18,Zunkovic18,Khasseh20}. For all-to-all symmetric interactions such as the LMG model, the thermodynamic limit ($N\rightarrow\infty$) of phase transitions is equivalent to the mean-field description~\cite{Munoz20}. This is seen in the classical phase space $(\theta,\phi)$ of the sphere, where the mean-field LMG is described by classical energy function,
\begin{align}
 E(s;\theta,\phi) &=\bra{\theta,\phi}\hat{H}(s,0)\ket{\theta,\phi}\nonumber\\ &=(s-1)\sin\theta\cos\phi-\frac{s}{2}\cos^2\theta,\label{eq:classE}
\end{align}
where $\ket{\theta,\phi}=e^{-i\phi\hat{J}_z}e^{-i\theta\hat{J}_y}\ket{\uparrow}^{\otimes N}$ is a spin-coherent state (SCS). For small $s$, this forms a single well centered on $(\theta,\phi)=(\frac{\pi}{2},0)$ as in \cref{fig:classicalPS}, but as $s$ grows to $s_{GQPT}=\frac{1}{2}$, this well undergoes a pitchfork bifurcation, splitting into a double-well with a separatrix between, and $(\frac{\pi}{2},0)$ becomes an unstable fixed point on the separatrix. Any point initialized below the separatrix energy will be bound to one well; any point above will orbit around both, shown in \cref{fig:classicalPS}. These two orbits can be distinguished in the thermodynamic limit by the order parameter of $Z$-polarization, $\langle J_z\rangle$. The ground-state quantum phase transition (GQPT) corresponds to spontaneous symmetry breaking of the ground state at the critical value $s_{GQPT}=1/2$, and so the GQPT is a kinematic property of the LMG's eigensystem. An order parameter for dynamical quantum phase transition (DQPT) corresponds to the long time-averaged of the Z-polarization, $\overline{\langle J_z(t)\rangle}$ after quench dynamics, with the state is initialized in a $Z$-polarized SCS~\cite{Mitra23}. In the thermodynamic limit, when $s=s_{DQPT}=2/3$, the initial state lies on the separatrix, corresponding to the DQPT critical point~\cite{Chinni22}.

Even without perfect permutation symmetry, $\hat{H}(s,\alpha)$ still obeys two other symmetries: global spin-flip parity and permutative mirror symmetries. That is, the global spin-flip operator $\exp(i\pi\sum_j\hat{\sigma}_x^{(j)})$ commutes with the Hamiltonian, and under the spin-site relabeling $J\to N+1-J$, the Hamiltonian is unchanged. Finally, the dynamics of $\hat{H}(s\approx 0,\alpha)$ is only perturbed slightly from simple Larmor precession generated by $\hat{J}_x$ for $s$ small. On the other hand, while $\hat{H}(s=1,\alpha)$ is a long-range one-axis-twisting model with integrable dynamics~\cite{Comparin22}. This integrability arises from severe degeneracies in the spectrum which immediately break upon the introduction of even a minuscule transverse field. For this reason, the chaos in $\hat{H}(s\approx 1,\alpha)$ does not resemble dynamics from $\hat{H}(s=1,\alpha)$. 

\subsection{Irreducible Representations}\label{sec:irreps}
\begin{figure}
\centering\includegraphics[width=0.95\linewidth]{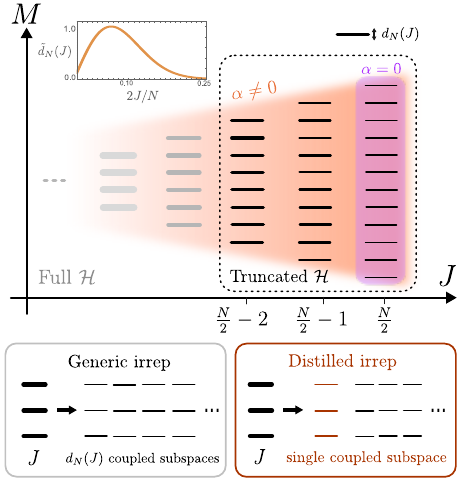} 
\caption{\label{fig:Irreps}(a) Schematic diagram of the Hilbert space structure for $N$ spin-$1/2$ particles. States of definite $\hat{J}^2$ (eigenvalues $J(J+1)$) and $\hat{J}_z$ (eigenvalues $M$) are shown. Infinite-range interactions in \cref{eq:Ham1}, corresponding to $\alpha=0$, can be described by states in the ``irrep'' with $J=N/2$, the symmetric subspace. When $\alpha\neq0$, the symmetric subspace becomes coupled to lower-$J$ subspaces, facilitating population-decay therefrom. The degeneracy of each irrep as a function of $J$ of \cref{eq:deg_dN} is depicted in the inset plot normalized by its maximum value, i.e. $\tilde{d}_N(J)=d_N(J)/d_N^{\mathrm{max}}$. While lower irreps are degenerate, under a change of basis that respects $(J,M)$ we can define nondegenerate ``distilled'' irreps that uniquely directly couple to the symmetric subspace.}
\end{figure}

As the interaction becomes finite but long-range ($0<\alpha\lesssim 2$) \cite{Pappalardi18}, we expect some preservation of collective dynamics. Central to our method, thus, is to organize Hilbert space according to a notion of the collective order, by leveraging irreducible representations of $SU(2)$. An irreducible representation (irrep) of $SU(2)$ is a non-zero representation that cannot be broken down into smaller, nontrivial subrepresentations. In standard parlance, an irrep can be used to denote a subspace of the vector space that is closed under ladder operators $\hat{J}_\pm\equiv \hat{J}_x\pm i\hat{J}_y$ with no closed subsets. We will use the term ``irrep" in this latter sence. Since the LMG Hamiltonian of \cref{eq:LMG} commutes with $\hat{J}^2$ (which is the Casimir element of $SU(2)$), its dynamics become closed within one such irrep determined by the initial state \cite{Arrechi72}. Formally, given $N$ spin-1/2 particles, the Hilbert space decomposes as
\begin{equation}
 \mathcal{H}= \mathcal{H}_{1/2}^{\otimes N}=\bigoplus_{J=(0,1/2)}^{N/2}\bigoplus_{u=1}^{d_N(J)}\mathcal{H}_{J,u}.\label{eq:CompBasis}
\end{equation}
The tensor product of reducible representations contrasts with the direct-sum structure of irreps. Each irrep is specified by the quantum number of $\hat{J}^2$ (denoted $J$), and another quantum number $u$ ranging from $u=1$ to 
\begin{equation}
 d_N(J) = \frac{2J+1}{N/2+J+1}\binom{N}{N/2-J}.
 \label{eq:deg_dN}
\end{equation}
Thus, there is a degenerate set of irreps with the same $J$. For general system size, the only nondegenerate irrep is when $J$ takes its maximum value, $J=N/2$, known as the ``symmetric subspace" as vectors in this irrep are symmetric under exchange of any two spin-halves when viewed as a tensor product state. The combinatoric nature of the degeneracy factor shows exponential growth for $J\ll N/2$, see inset in \cref{fig:Irreps}.

The standard basis of the irreps is the Dicke basis~\cite{Dicke54}, the simultaneous eigenvectors of $\hat{J}^2$ and $\hat{J}_z$, augmented by the degeneracy $\ket{J,M,u}$. Local degrees of freedom in the separable basis in \cref{eq:CompBasis}, $\mathcal{H}_{1/2}=\text{span}(\ket{\uparrow},\ket{\downarrow})$, map onto qubits, $\{\ket{\uparrow},\ket{\downarrow}\}\cong\{\ket{0},\ket{1}\}$ and in the context of quantum computing, their tensor product is called the ``computational basis." An arbitrary computational state is represented $\ket{\Vec{s}}\equiv\ket{s_1}\otimes\ket{s_2}\otimes...\ket{s_N}$ for $s_j\in\{0,1\}$.

The canonical change of basis from the Dicke basis to the computational basis is a recursive algorithm; there is no \textit{closed-form} linear map between the two. The inner products $\braket{\Vec{s}}{J,M,u}$ come from the recursion relation of the Clebsch-Gordan series~\cite{Arrechi72},
\begin{align}
 \ket{J,M,u}=&C^{J,M}_{J-\frac{1}{2},M-\frac{1}{2};\frac{1}{2},\frac{1}{2}}\ket{\uparrow}\ket{J-\frac{1}{2},M-\frac{1}{2}}+&\nonumber
 \\&C^{J,M}_{J-\frac{1}{2},M+\frac{1}{2};\frac{1}{2},-\frac{1}{2}}\ket{\downarrow}\ket{J-\frac{1}{2},M+\frac{1}{2}},&\label{eq:CGSeries1}\\
 \ket{J,M,v}=&C^{J,M}_{J+\frac{1}{2},M-\frac{1}{2};\frac{1}{2},\frac{1}{2}}\ket{\uparrow}\ket{J+\frac{1}{2},M-\frac{1}{2}}+&\nonumber
 \\&C^{J,M}_{J+\frac{1}{2},M+\frac{1}{2};\frac{1}{2},-\frac{1}{2}}\ket{\downarrow}\ket{J+\frac{1}{2},M+\frac{1}{2}},&\label{eq:CGSeries2}
\end{align}
where $C^{J,M}_{j_1,m_1;j_2,m_2}=\langle J,M | j_1,m_1;j_2,m_2 \rangle$ is a Clebsch-Gordan coefficient~\cite{Biedenharn81}. The procedure reveals that irrep degeneracy emerges from the choice of adding or subtracting subsystem angular momenta to construct total angular momentum. Beginning from a single degree of freedom $J=\frac{1}{2}$, the recursion proceeds by attaching another single spin, and using the Clebsch-Gordan series to split the levels into an additive $J+\frac{1}{2}$ and subtractive $J-\frac{1}{2}$ subspace. Attaching the next single spin, each subspace undergoes another round of level-splitting into additive and subtractive forms, et cetera. This recursive method becomes tedious and intractable as system size increases, especially for lower irreps $J \ll \frac{N}{2}$, where the degeneracy is most severe. Moreover, since $\ket{J,M,u}$ behaves equivalently for all $u$ under symmetric operations, the index $u$ is arbitrary and its derivation from Clebsch-Gordan recursion is unnecessary for our purposes.

Instead, we seek a nonrecursive mapping between computational and Dicke states, beginning from the symmetric subspace. One can achieve a ``top-down" change of basis transformation between the computational and irrep basis as follows. The symmetric subspace $\mathcal{H}_{N/2}$ is spanned by the Dicke States,
\begin{equation}
 \ket{\frac{N}{2},M}=\{\ket{\uparrow}^{\otimes\frac{N}{2}+M}\ket{\downarrow}^{\otimes\frac{N}{2}-M}\}_{\mathrm{sym}},
\end{equation}
which are permutation-symmetric superpositions of $\ket{\Vec{s}}$ with $\sum_j s_j=\frac{N}{2}-M$. More generally, one may define Dicke states from stretch states under the action of ladder operators,
\begin{equation}
 \ket{J,M,u}=\frac{\hat{J}_-^{J-M}\ket{J,J,u}}{||\hat{J}_-^{J-M}\ket{J,J,u}||},\label{eq:ladder}
\end{equation}
directly from the definition of an irrep. The first irreps below the symmetric subspace, $\mathcal{H}_{N/2-1,u}$, are spanned by ladder operations (\cref{eq:ladder}) on each irrep's stretch state
\begin{equation}
 \ket{\frac{N}{2}-1,\frac{N}{2}-1,u}=\sum_j c^{(1,u)}_j\ket{\uparrow}^{\otimes j-1}\ket{\downarrow}_j\ket{\uparrow}^{\otimes N-j},\label{eq:1down}
\end{equation}
where $\ket{\uparrow}^{\otimes j-1}\ket{\downarrow}_j\ket{\uparrow}^{\otimes N-j}$ will be abbreviated $\ket{\downarrow_j}$ henceforth. Tautologically, $c^{(1,u)}_j\equiv \braket{\downarrow_j}{\frac{N}{2}-1,\frac{N}{2}-1,u}$ are ``computational amplitudes," but are nonunique whenever the first irreps are degenerate with multiple possible values of quantum number $u$. Similarly, within the second irreps $\mathcal{H}_{N/2-2,u}$, the stretch states are $\ket{\frac{N}{2}-2,\frac{N}{2}-2,u}=\sum_{j>k}c^{(2,u)}_{j,k}\ket{\downarrow_j\downarrow_k}$. Valid computational amplitudes must satisfy the following orthonormality relations:
\begin{align}
 \sum_j c^{(1,u)}_j=0=\sum_j c^{(2,u)}_{j,k},\label{eq:compamp1}\\
 \sum_j c^{(1,u)}_j c^{(1,v)*}_j=\delta_{uv}=\frac{1}{2}\sum_{j,k} c^{(2,u)}_{j,k} c^{(2,v)*}_{j,k},\label{eq:compamp2}\\
 c^{(2,u)}_{j,k}=c^{(2,u)}_{k,j}.
 \label{eq:compamp3}
\end{align}
\Cref{eq:compamp1,eq:compamp2,eq:compamp3} define the computational amplitudes. Stretched states of an irrep are annihilated by ladder operators $\hat{J}_+$ and the set of degeneracy indexed computational amplitudes form an orthonormal basis. \Cref{eq:compamp3} follows from the commutativity of every spin-site degree of freedom with every other, since $\hat{\sigma}_z^{(j)}\hat{\sigma}_z^{(k)}=\hat{\sigma}_z^{(k)}\hat{\sigma}_z^{(j)}$. These constraints capaciously and specifically generate states with quantum numbers $J\in\{\frac{N}{2}-1,\frac{N}{2}-2\}$ and $M=J$, and so map onto the Dicke states obtained from Clebsch-Gordan recursion in \cref{eq:CGSeries1,eq:CGSeries2} via some unitary change-of-basis. We see then the freedom in defining the irreps within the degenerate manifolds. This freedom will be central to our method for ``distilling'' irreps to most efficiently approximate the dynamics of long-range interacting systems.

For time-independent symmetric all-to-all coupling models, such as the LMG, the Hamiltonian commutes with $\hat{J}^2$, and thus is block diagonal in the Dicke basis, with each irrep a closed block. A state initialized in the symmetric subspace will remain dynamically therein and restricted to an $N+1$-dimensional Hilbert space, not $2^N$-dimensional. This block-closure of Hilbert space permits exact diagonalization of the LMG at general system sizes~\cite{Debergh01,Morita06}. In contrast, the long-range Ising Hamiltonian generically couples all irreps dynamically. When the range exponent $\alpha>0$, the collective spin symmetry is broken, and it has been shown that immediately the Hamiltonian becomes chaotic, and satisfies the ETH over the bulk of its spectrum~\cite{Lerose23}. For very long-range models, and for some parameters $s$, however, full ergodic exploration of the full exponentially large Hilbert space does not occur. In these cases an $N+1$-dimensional subset of eigenstates of $\hat{H}(s,\alpha)$ violates the ETH, which are the quantum many-body scars (QMBS). The scar subspace $\mathcal{H}_{QMBS}$ maps to the symmetric subspace $\mathcal{H}_{J=N/2}$, such that $\lim_{\alpha\to 0}\mathcal{H}_{QMBS}=\mathcal{H}_{N/2}$. For this reason, initializing a system governed by $\hat{H}(s,\alpha)$ in a symmetric state $\ket{\psi_0}\in\mathcal{H}_{N/2}$ should result in dynamics dominated by the QMBS, and consequently remaining mostly within and close to the symmetric subspace. We will test this hypothesis by constructing the general matrix elements of the transverse Ising Hamiltonian in the basis of Dicke states, up to the indeterminacy of irreps. We will then truncate Hilbert space to keep only the states to which the symmetric subspace is directly coupled by the Hamiltonian's off-diagonal matrix-elements and study range of validity of the resulting approximation.

\subsection{Hamiltonian Matrix Elements}\label{sec:HME}
The permutation-asymmetric operator in $\hat{H}(s,\alpha)$ contains weight-2 interactions $\hat{\sigma}_z^{(j)}\hat{\sigma}_z^{(k)}$, and thus the Hamiltonian directly couples irreps differing by no more than 2 in total angular momenta,
\begin{equation}
 \bra{J',M'}\hat{H}(s,\alpha)\ket{J,M}\neq 0\quad \implies\quad |J-J'|\leq 2,
 \label{eq:OffDiagH}
\end{equation}
due to selection-rules. This implies that to first order, the perturbed eigenstates will contain a small admixture of irreps $\mathcal{H}_{N/2-1,u}$ and $\mathcal{H}_{N/2-2,u}$, and no other subspaces couple to the symmetric subspace at this lowest order. These high-$J$ irreps are also the simplest to express, since they represent few-body excitations of otherwise symmetric collective states.

We seek a decomposition of $\hat{H}(s,\alpha)$ in the irrep basis of Dicke states. This requires an operator basis of irreducible operators to replace the \textit{reducible} operator basis of Pauli products, $\hat{P}_{\alpha_1,\alpha_2,\dots, \alpha_N} = \bigotimes_{i=1}^N \hat{\sigma}_{\alpha_i}$ (including $\hat{\sigma}_0\equiv \hat{I}$). These operators in the basis we seek should also transform as tensors under $SU(2)$, to guarantee \cref{eq:OffDiagH}. These criteria are satisfied by the generalized irreducible spherical tensor operators that map between irreps $\hat{T}^{(k)}_q:\mathcal{H}_{J',v}\to\mathcal{H}_{J,u}$, which obey the Wigner-Eckart theorem, defined as~\cite{Klimov08,Klimov17}

\begin{align}
&\hat{T}^{(k)}_q[J,u;J',v]\equiv\label{eq:sphT}\nonumber\\
&\sqrt{\frac{2k+1}{2J+1}}\sum_{M=-J'}^{J'} 
C^{J,M+q}_{J',M;k,q}\ket{J,M+q,u}\bra{J',M,v}.
\end{align}
The generalized spherical tensor operators satisfy the requirements in \cref{eq:OffDiagH}, 
\begin{equation}
 \bra{J',M'}\hat{T}^{(k)}_q\ket{J,M}\neq 0\quad \implies\quad |J-J'|\leq k,
 \label{eq:OffDiagT}
\end{equation}
due to Clebsch-Gordan selection rules. All tensor operators that map a given irrep to itself, $(J,u)=(J',v)$ are proportional to solid harmonics of angular momenta, projected onto that irrep,
\begin{equation}
 \hat{T}^{(k)}_q[J,u;J,u]\propto \hat{\Pi}_{J,u}\mathcal{Y}^k_q(\hat{\mathbf{J}}),
\end{equation}
where the projectors are $\hat{\Pi}_{J,u}\equiv \sum_M\ket{J,M,u}\bra{J,M,u}$ and the solid harmonics are $\mathcal{Y}^{(k)}_q(r,\theta,\phi)\equiv \sqrt{\frac{4\pi}{2k+1}}r^k Y^{(k)}_q(\theta,\phi)$ derived from spherical harmonics $Y^{(k)}_q$. Otherwise, when coupling two different irreps, spherical tensors cannot decompose into collective operators, but remain proportional to $q$-projected $k$-order products of Pauli operators. The generalized spherical tensors thus give a parsimonious repesentation of the Hamiltonian, as $\hat{H}(s,\alpha)\propto\hat{T}^{(k)}_q$ only for $k\leq 2$.

Pauli strings $\hat{P}_{\alpha_1,\alpha_2,\dots, \alpha_N}$ and generalized spherical tensor operators $\hat{T}^{(k)}_q[J,u;J',v]$ both form complete orthonormal bases for operators on the $N$-qubit Hilbert space. The former is natural when considering local dynamics where the latter is natural for considering collective dynamics, especially when states occupy the symmetric subspace and topmost irreps. To express the Hamiltonian in the spherical tensor basis, note that
$\hat{\sigma}_x^{(j)}$ is spanned by $\{\hat{T}^{(1)}_1[J,u;J',v], \; \hat{T}^{(1)}_{-1}[J,u;J',v]\}$ and $\hat{\sigma}_z^{(j)}\hat{\sigma}_z^{(k)}$ is spanned by$\{\hat{T}^{(0)}_0[J,u;J,v], \;\hat{T}^{(2)}_0[J,u;J',v]\}$. Furthermore, the uniform transverse field is permutation symmetric and thus does not couple different irreps, so we can express $\hat{H}(s,\alpha)$ in the spherical tensor basis as
\begin{align}
\hat{H}(s,\alpha)=s\sum_{J,u,v}F_0[\alpha,J,u,v]\hat{T}^{(0)}_0[J,u;J,v]\nonumber\\
+(s-1)\sum_{J,u}F_1[J](\hat{T}^{(1)}_1[J,u;J,u]-\hat{T}^{(1)}
_{-1}[J,u;J,u])\nonumber\\
+s\sum_{J,u,v}\sum_{\delta_J=-2}^{2}F_2[\alpha,J,u;J+\delta_J,v]\hat{T}^{(2)}_0[J,u;J+\delta_J,v],\label{eq:Ham2}
\end{align}
a comparatively simple tensor structure, with no participation by $\hat{T}^{(k>2)}_q$. The scalar weights $F_0,F_1,F_2$ on each spherical tensor arise from their mutual trace with Pauli operators,
\begin{align}
 &F_0[\alpha,J,u,v]=&\nonumber\\ 
 &\frac{-1}{4\mathcal{N}}\sum_{j,k}|j-k|^{-\alpha}\Tr(\hat{\sigma}_z^{(j)}\hat{\sigma}_z^{(k)}\hat{T}^{(0)}_0[J,u;J,v]),&\label{eq:F0}\\
 &F_1[J]=&\nonumber\\
 &\frac{1}{2}\sum_j \Tr(\hat{\sigma}_x^{(j)}\hat{T}^{(1)}_1[J,u;J,u]),&\label{eq:F1}\\
 &F_2[\alpha,J,u;J',v]=&\nonumber\\ 
 &\frac{-1}{4\mathcal{N}}\sum_{j,k}|j-k|^{-\alpha}\Tr(\hat{\sigma}_z^{(j)}\hat{\sigma}_z^{(k)}\hat{T}^{(2)}_0[J,u;J',v]).&\label{eq:F2}
\end{align}
While $F_1$ affords closed-form solutions by simple Clebsch-Gordan rules, $F_0$ and $F_2$ are calculated by analyzing the Dicke-basis matrix elements $\bra{J,M,u}\hat{\sigma}_z^{(j)}\hat{\sigma}_z^{(k)}\ket{J',M,v}$ in order to solve the mutual traces. The complete derivation is given in \cref{app:Fsol}. The most salient weights are those that couple $\mathcal{H}_{N/2}$ to $\mathcal{H}_{N/2-1,u}$ and $\mathcal{H}_{N/2-2,u}$,
\begin{align}
 &F_2[\alpha,\frac{N}{2},0;\frac{N}{2}-1,u]\nonumber\\
 &=\frac{-1}{2\mathcal{N}}\sqrt{\frac{(N+2)(N+1)}{30(N-1)}}\sum_{j,k}\frac{c^{(1,u)}_j+c^{(1,u)}_k}{|j-k|^{\alpha}},\label{eq:F2co1}\\
 &F_2[\alpha,\frac{N}{2},0;\frac{N}{2}-2,u]\nonumber\\
 &=\frac{-1}{\mathcal{N}}\sqrt{\frac{N+1}{30}}\sum_{j,k}\frac{c^{(2,u)}_{jk}}{|j-k|^{\alpha}},\label{eq:F2co2}
\end{align}
which remain undefined until the computational amplitudes for each irrep are determined in accordance with the orthonormality rules \cref{eq:compamp1,eq:compamp2,eq:compamp3}.

\subsection{Irrep Distillation}\label{sec:IRD}
As discussed in Sec. \ref{sec:irreps}, there is freedom in the choice of computational basis amplitudes due to the degeneracy of the irreps for $J<N/2$ within the constraints of \cref{eq:compamp1,eq:compamp2,eq:compamp3}. This permits optimization of the sums $\sum_{j,k}|j-k|^{-\alpha}(c^{(1,u)}_j+c^{(1,u)}_k)$ and $\sum_{j,k}|j-k|^{-\alpha}c^{(2,u)}_{j,k}$ in \cref{eq:F2co1,eq:F2co2} respectively. In particular, we can choose our irreps within the degenerate subspace to align most efficiently with the the dynamic generated by the Hamiltonian. We achieve maximum magnitude in each of these sums with ``bright" irreps which maximally couple to the symmetric subspace via the Hamiltonian, in a method we call irreducible representation distillation (IRD). In fact this coupling is kinematically exclusive; all other irreps are accessible from the symmetric subspace only by \textit{multiple} actions of the Hamiltonian. Thus IRD solves the minimal dimension Hilbert space to first order perturbations beyond the symmetric subspace. 

\begin{figure}[t]
\centering\includegraphics[width=\linewidth]{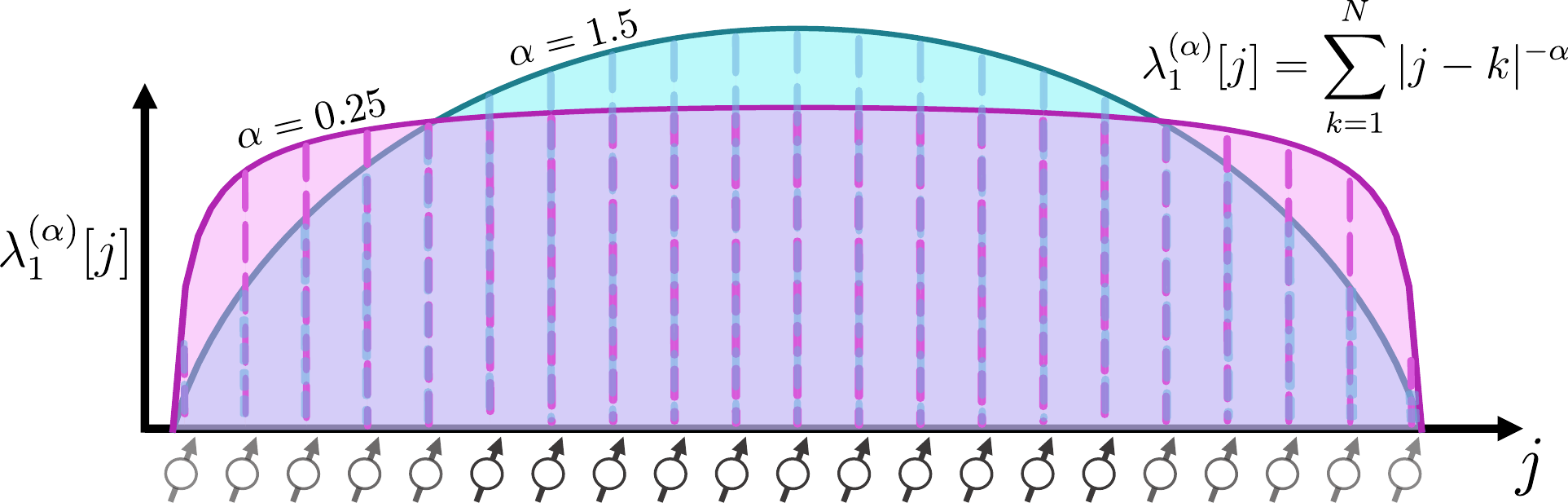}
\caption{\label{fig:distilled_irrep_illustration}Illustration of magnitudes of the amplitude on the $\ket{\downarrow_j}$ state of the distilled irrep corresponding to $J=N/2-1$ as in \cref{eq:1down}. Here the ``localizing function" weight $\lambda_1^{(\alpha)}[j]$ is proportional to $c_j^{(1,1)}$ in \cref{eq:dist1} up to norm and a constant offset. Shown are two values of $\alpha$, for $\alpha=0.25$ in magenta and $\alpha=1.5$ in cyan. This choice of amplitudes ensures that the irrep constructed maximizes population transfer from the symmetric subspace for early timescales, hence it is \textit{distilled}.}
\end{figure}

For the distilled irreps, the computational amplitudes are given by sums of the ``localizing function" across one or more of its (symmetric) arguments: $\lambda^{(\alpha)}_1[j]\equiv \sum_{k=1}^N |j-k|^{-\alpha}$ and $\lambda^{(\alpha)}_0\equiv \sum_{j=1}^N\sum_{k=1}^{j-1} |j-k|^{-\alpha}$. From these sums, which marginalize the localizing function $|j-k|^{-\alpha}$ as shown in \cref{fig:distilled_irrep_illustration}, the distilled computational amplitudes $\Vec{c}^{(1,1)}, \Vec{c}^{(2,1)}$ are then made to satisfy \cref{eq:compamp1,eq:compamp2,eq:compamp3} as

\begin{align}
&c^{(1,1)}_j=\nonumber\\
&\frac{\lambda^{(\alpha)}_1[j] -\frac{2}{N}\lambda^{(\alpha)}_0}{\sqrt{(\sum_j \lambda^{(\alpha)2}_1[j])- \frac{4}{N}\lambda^{(\alpha)2}_0}},\label{eq:dist1}\\
&c^{(2,1)}_{j,k}=\nonumber\\
&\frac{(N-2)|j-k|^{-\alpha}-\lambda^{(\alpha)}_1[j]-\lambda^{(\alpha)}_1[k]+\frac{2}{N-1}\lambda^{(\alpha)}_0}{\sqrt{N-2}\sqrt{(N-2)\lambda^{(2\alpha)}_0+\frac{2}{N-1}\lambda^{(\alpha)2}_0-\sum_j \lambda^{(\alpha)2}_1[j]}}.\label{eq:dist2}
\end{align}
It comes as no coincidence that distilled computational amplitudes take the form of orthonormalized marginals of the localizing function, because the sums in \cref{eq:F2co1,eq:F2co2} describe an inner product between the localizing function and the computational amplitudes in an $\binom{N}{1}$- or $\binom{N}{2}$-dimensional vector space. As a consequence of this maximum and the prior orthonormality relations, 
\begin{align}
 \sum_j c^{(1,1)}_jc^{(1,u)*}_j=0&\implies \sum_{j,k}\frac{c^{(1,u)}_{j}+c^{(1,u)}_k}{|j-k|^{\alpha}}=0,\\
 \sum_{j,k} c^{(2,1)}_{j,k}c^{(2,u)*}_{j,k}=0&\implies \sum_{j,k}\frac{c^{(2,u)}_{j,k}}{|j-k|^{\alpha}}=0,
\end{align}
meaning that to lowest order in perturbation theory, the Hamiltonian couples exactly three irreps to the symmetric subspace: the distilled irreps $\mathcal{H}_{N/2-1,1}$ and $\mathcal{H}_{N/2-2,1}$, and the symmetric subspace $\mathcal{H}_{N/2}$ itself. All non-distilled irreps with $J\geq \frac{N}{2}-2$, as well as all irreps with $J<\frac{N}{2}-2$, only affect dynamics at higher perturbative orders. We benefit from IRD by acquiring sufficient matrix elements to treat the QMBS as a first-order perturbation theory on LMG eigenstates. To this order, dynamics on exponential Hilbert space is restricted to the distilled subspace,
\begin{equation}
\mathcal{H}_D\equiv\mathcal{H}_{N/2}\oplus\mathcal{H}_{N/2-1,1}\oplus\mathcal{H}_{N/2-2,1}, \label{eq:calHD}
\end{equation}
with $||\mathcal{H}_D||=3(N-1)$. This method of IRD generalizes to permutation symmetry-breaking functions beyond the algebraic decay function considered here, $|j-k|^{-\alpha}$, to any non-constant coefficient with non-constant marginals. Upon transforming into these distilled irreps and truncating Hilbert space down to $\mathcal{H}_D$, we obtain an approximation of the Hamiltonian $\hat{H}_D(s,\alpha):\mathcal{H}_D\to\mathcal{H}_D$, whose properties we discuss below.

\section{Applications}\label{sec:app}
\subsection{Spectra and QMBS}\label{sec:spec}
Numerical comparisons between the distilled and exact systems are limited by the dimension of the exact Hamiltonian, which scales as $2^N$. We losslessly compress this dimension by exploiting the aforementioned spin-flip and mirror symmetries in the Hamiltonian to generate only the positive eigenstates of both symmetry operators when diagonalizing. This slows, but does not eliminate the exponential scaling of Hilbert space. 

\begin{figure*}[ht]
\centering
\includegraphics[width=\linewidth]{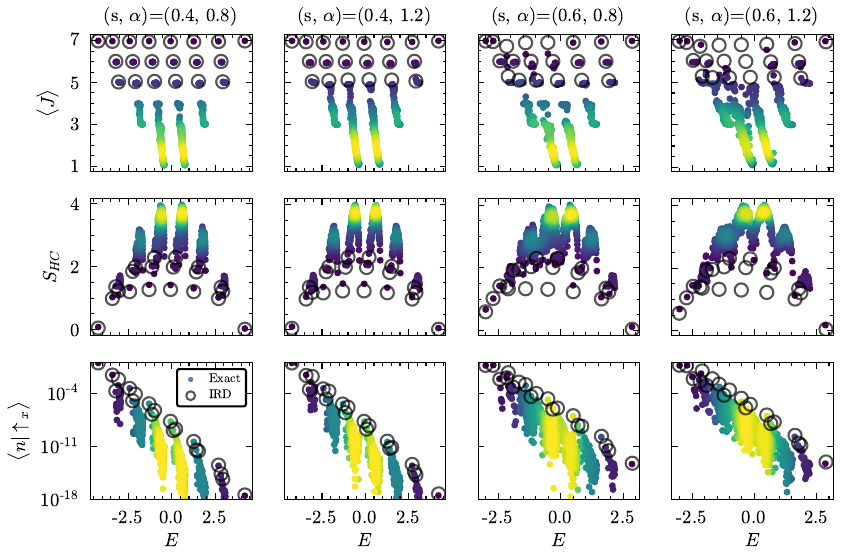}
\caption{\label{fig:DistSpec} Eigenstate properties of the exact Hamiltonian, $\hat{H}(s,\alpha)$, compared with the distilled and truncated Hamiltonian, $\hat{H}_D(s,\alpha)$. For eigenstates $\ket{n}$ (colored dots) and $\ket{n_D}$ (empty circles), plots in each row show spin-number $\langle J\rangle$, half-chain entanglement entropy $S_{HC}$, and overlap with $X$-polarized SCS, $|\braket{n}{J=\frac{N}{2},M_x=\frac{N}{2}}|^2$. Two left-most columns with $s=0.4$ show full QMBS across 8 eigenstates, two right-most columns with $s=0.6$ show broken QMBS. Dot color shows the density of data points due to irrep degeneracy, with yellow representing dense distributions.}
\end{figure*}

\Cref{fig:DistSpec} introduces the quantum many body scars of $\hat{H}(s,\alpha)$ by plotting some of these eigenstates' notable properties compared with the rest of the eigensystem. The QMBS in the long-range transverse Ising model are a set of $N+1$ eigenstates, related to the Anderson tower of states~\cite{Comparin22}, spanning the spectrum with regular, smoothly varying energy separations~\cite{Lerose23}, including the ground state and maximum energy states. Whereas most eigenstates have similar properties to those in their energy neighborhoods~\cite{Russomanno21}, the QMBS, as shown in \cref{fig:DistSpec}, show much more symmetric properties than their energy neighbors. The QMBS are the eigenstates with high population in the symmetric subspace $\langle \hat{\Pi}_{N/2}\rangle >\frac{1}{2}$. Consequently, they take high spin-numbers $\langle J\rangle \gtrsim \frac{N-1}{2}$, calculated as the expectation of $\sum_{J,u}J\langle\hat{\Pi}_{J,u}\rangle$, high overlap with spin-coherent states $|\braket{n_{QMBS}}{\theta,\phi}|^2\gg 0$, and low entanglement entropy. The QMBS are the top $\langle J\rangle$-row of exact eigenstates in \cref{fig:DistSpec}(a-b), and for every scar state there is a corresponding approximate scar state in the eigensystem of the distilled Hamiltonian $\hat{H}_D$. However, non-scar eigenstates of $\hat{H}_D$ (the second and third rows in $\langle J\rangle$) do not have high overlap with any exact eigenstates. \Cref{fig:DistSpec} demonstrates not only examples of the spectral form of the Hamiltonian at $N=14$, but a few places where it is known to form QMBS, e.g., at $(s,\alpha)=(0.4,0.8\sim 1.2)$, and where the QMBS is broken, such as at $(s,\alpha)=(0.6,0.8\sim 1.2)$. A ``broken" QMBS is one which does not cover the full rank of the symmetric subspace. Instead, some eigenstates have significant population in both $\mathcal{H}_{N/2}$ and other irreps; these are the hybridized eigenstates and thus not scars. The close agreement of distilled eigenstates $\ket{n_D}$ to their exact counterparts $\ket{n_{QMBS}}$ in terms of these observables indicates only that the approximation captures certain macroscopic properties and does not on its own clarify the approximation fidelity $|\braket{n_D}{n_{QMBS}}|^2$ nor the bounds on the QMBS regime. 

\begin{figure}
\centering
\includegraphics[width=\linewidth]{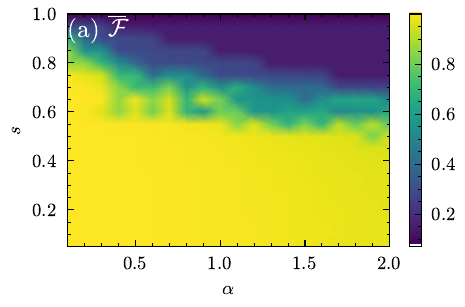}
\includegraphics[width=\linewidth]{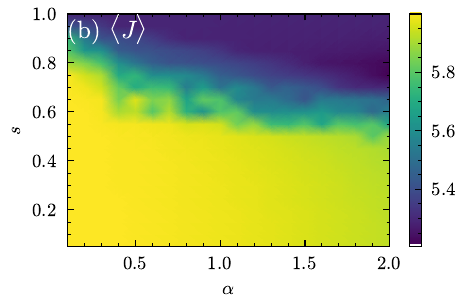}
\caption{\label{fig:Str} Two different measures of the QMBS, for $N=12$ across $(s,\alpha)$. (a) average fidelity $\overline{\mathcal{F}}$ for $\mathcal{F}=|\braket{n_{QMBS}}{n_{D}}|^2$ between the exact and distilled spectra. (b) spin-number $\overline{\langle J\rangle}$ over the same regime. The parameters in which the spectrum shows the strongest QMBS are also those in which distillation best captures these eigenstates, and this small system foreshadows the QMBS regime of larger systems: guaranteed scars for $s\lesssim0.42$, though the small system is more forgiving of large $\alpha$.}
\end{figure}

Under distillation and truncation, the spectra of the resulting Hamiltonian successfully reproduces QMBS for all parameters where QMBS appears in the exact Hamiltonian. Due to finite-size effects at small $N$, distillation also sometimes reconstructs false-QMBS even with parameters at which the exact Hamiltonian is not scarred. Insofar as the QMBS consists of eigenstates whose dominant population is in $\mathcal{H}_{N/2}$, exhibiting spin-numbers $\langle J\rangle\approx \frac{N}{2}$, couplings between the symmetric subspace and lower irreps is weak, justifying truncation. We proceed to measure ensemble-average properties over the ensemble of eigenstates with $\langle \Pi_{N/2}\rangle\geq \frac{1}{2}$, which are the candidates for QMBS. Under this logic, we expect that for a given scar state $\ket{n_{QMBS}}$, there should be close tracking between $\langle J\rangle$ and IRD fidelity $\mathcal{F}\equiv |\braket{n_{QMBS}}{n_D}|^2$ despite there being no intrinsic connection between the two metrics. Consulting \cref{fig:Str}, we see exactly this tracking: the ensemble-average fidelity is high ($\overline{\mathcal{F}}\approx 1$) in \cref{fig:Str}a exactly where the ensemble-average spin-number is near-maximum ($\overline{\langle J\rangle}\approx \frac{N}{2}$) in \cref{fig:Str}b. We see unit-average fidelities between distilled and exact scar states at every tractable system size, within well-defined parametric bounds. We will show in Sec. \ref{sec:delE} that these bounds grow stricter with increasing $N$, but are depicted at the finite size of $N=12$ in \cref{fig:Str}. There appears a broad region of strong QMBS for $s<0.6$ and all $\alpha$, but with some slow decay with growing $\alpha>1$, and a liminal region of QMBS for $0.6<s<0.8$ and $\alpha\ll 1$. The more permissive preservation of QMBS at extremely long-range, and slowly decaying QMBS as interaction ranges shrink, is precedented in the literature. Of more interest to us is the sharp boundary between QMBS and pure chaos brought about by small changes in $s$ when the driving and interaction terms in the Hamiltonian are on the same scale. This suggests that the existence of QMBS as a function of $s$ may follow a phase transition.

Breaches of the QMBS differ according to $(s,\alpha)$. At finite (though large) system size, shortening the range of interaction by increasing $\alpha$ gradually subverts scar states in the bulk of the spectrum due to strong coupling with lower irreps. Insufficient transverse driving (large $s$) allows spontaneous collapse of a few scar states due to strong hybridization with lower-irrep states in the energy neighborhood, as explained in \cite{Lerose23}. These degeneracies begin in the low-energy region of the spectrum near the ground state, and progress to higher energies as $s$ grows beyond the critical point. The breach in the QMBS tracks with the unstable fixed point resulting from the classical energy function's bifurcation discussed in Sec. \ref{sec:model}, which splits the single-well potential into two by raising the center, as associate with the GQPT. We hypothesize, therefore, that the instability seen in the classical phase space is associated with spontaneous violations of the QMBS. In the next section we show how perturbation theory justifies the relationship between the quality of approximation by IRD and the existence of QMBS, by showing that the region of distillation self-consistency exactly matches the QMBS regime.

\subsection{Bounds on QMBS}\label{sec:delE}
Upon truncating to the distilled irreps, the remaining Hilbert space is $3(N-1)$-dimensional, far smaller than the full $2^N$-dimensional Hilbert space, and reproduces the QMBS up to first-order perturbative expansions in energy eigenstates. Consider a perturbation theory where the zeroth order is the LMG Hamiltonian from \cref{eq:LMG} and the perturbation is $\hat{V}(s,\alpha)$. The eigenstates in the symmetric subspace, $\ket{n_{\text{sym}}}$, are those obeying \cref{eq:LMGs1,eq:LMGs2} with $\hat{J}^2\ket{n_{\text{sym}}}=\frac{N}{2}(\frac{N}{2}+1)\ket{n_{\text{sym}}}$ and thus $\ket{n_{\text{sym}}}\in\mathcal{H}_{N/2}$. The perturbed eigenenergies and eigenstates in this subspace are
\begin{align}
 E_n'&=E_n+\bra{n_{\text{sym}}}\hat{V}(\alpha)\ket{n_{\text{sym}}},\label{eq:pert1}\\
 \ket{n'}&=\ket{n_{\text{sym}}}+\sum_{m\neq n}\frac{\bra{m}\hat{V}(\alpha)\ket{n_{\text{sym}}}}{E_n-E_m}\ket{m},\label{eq:pert2}
\end{align}
where $\ket{m}$ runs over \textit{all} eigenstates of $\hat{H}(s,0)$ in every irrep, not only those in the symmetric subspace. Whether one defines the perturbation $\hat{V}(\alpha)\equiv \hat{H}(s,\alpha)-\hat{H}(s,0)$ or $\hat{V}(\alpha)\equiv \hat{H}_D(s,\alpha)-\hat{H}_D(s,0)$ where $\hat{H}_D$ is the irrep-distilled, truncated Hamiltonian, the first-order perturbed eigensystem in \cref{eq:pert1,eq:pert2} is unchanged. This is because the above truncation does not affect any matrix elements mapping to or from $\mathcal{H}_{N/2}$. Furthermore, because the QMBS are the eigenstates of $\hat{H}(s,\alpha)$ which are ``dressed" symmetric eigenstates, a perturbation theory of the dressed symmetric eigenstates is also one of the QMBS.

\begin{figure}
\centering
\includegraphics[width=\linewidth]{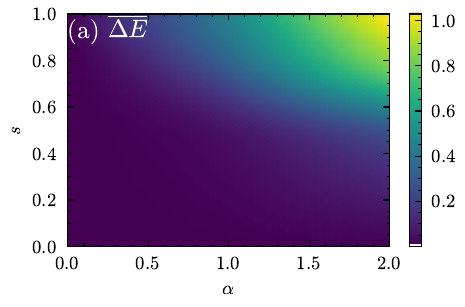}
\includegraphics[width=\linewidth]{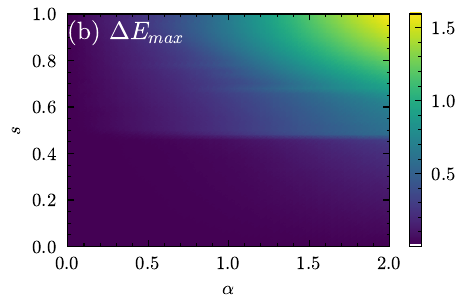}
\caption{\label{fig:varEParam} Energy uncertainty $\Delta E$, defined in \cref{eq:energy_variance}, as a measure of the accuracy of the distilled eigenstates for $N=100$, across $(s,\alpha)$. (a) scar-average variance $\overline{\Delta E}$. (b) scar-maximum variance $\Delta E_{max}$. All energy variances remain small when $s< 0.5$ at this system size. The maximum $\Delta E_{max}$ shows a sharp jump at $s=0.5$.}
\end{figure}

The degree to which the perturbed eigenstates are not eigenstates of the exact Hamiltonian can be quantified by their energy uncertainty with respect the desired Hamiltonian, 
\begin{equation}
 \Delta E_{n'}\equiv \sqrt{\bra{n'}\hat{H}_D^2\ket{n'}-\bra{n'}\hat{H}_D\ket{n'}^2}.
 \label{eq:energy_variance}
\end{equation}
This uncertainty forms the leading order correction to eigenstate evolution, and consequently dictates the rate of short-time decay in dynamical fidelity, such as Loschmidt echoes (to be described in Sec. \ref{sec:NCS}). The growth of energy variance, especially of its ensemble average $\overline{\Delta E}\equiv \sum_{n'}\frac{\Delta E_{n'}}{N+1}$, and ensemble maximum $\Delta E_{max}\equiv \max_{n'}\Delta E_{n'}$, predict worst-case performance of the convergence of perturbation theory, and consequently of the QMBS itself. We use $\Delta E_{max}$ as a self-consistency test of IRD at large system-size.

\Cref{fig:varEParam} depicts the growth of energy variance in the perturbed eigenstates as $(s,\alpha)$ increase. The sharp increase in $\Delta E_{max}$ at $s=0.5$ in \cref{fig:varEParam}b coincides with the emergence of an unstable fixed point in $E(s;\theta,\phi)$ from \cref{eq:classE} which gives rise to the GQPT. Finite-size scaling analysis shows the placement of this discontinuity at $s=0.5$ is a finite-size effect, and in the thermodynamic limit the critical point is $s_c\approx 0.42$, as we find in \Cref{app:FSSA}. By contrast, \cref{fig:varEParam}a is smooth, $\overline{\Delta E}$ is insensitive to the critical point. This indicates that the breakdown in the QMBS must start locally, affecting one or a few eigenstates, before globalizing across the QMBS. In any case, when $\Delta E$ reaches a significant fraction of the inter-scar level spacing, then perturbation theory and IRD no longer agree or converge, and the dynamical fidelity (measured by Loschmidt echoes) of the non-convergent scar state decays rapidly. A similar, though more gradual boundary at $\alpha=1$ (also observed in Loschmidt echoes) limits the long-range regime to the lattice dimension, corroborating previous literature, and this boundary likewise correlates with the preservation or decay of dynamical fidelity.

Notably, the bounds on IRD self-consistency indicated by \cref{fig:varEParam} align with the parameters where QMBS is known to exist at finite system size, attested to by \cref{fig:Str}. Thus, not only can IRD capture the spectral and state-properties of QMBS within the QMBS regime, it can test its own self-consistency without recourse to the full-dimensional system, and is self-consistent wherever there exists an unbroken QMBS.

\subsection{Quantum Phase Transitions}\label{sec:QPT}
Having demonstrated that IRD-truncation well-approximates the QMBS, we now consider whether distillation preserves order parameters in long-range interacting spin Hamiltonians \textit{outside} the QMBS regime, such as QPTs. The GQPT and DQPT defined in Sec. \ref{sec:model}, both are defined by an order parameter given by the system polarization $\langle J_z\rangle$. Since $\langle J_z\rangle$ is a ``macro" observable, two dynamical systems whose microstates severely differ may still agree on that macrostate~\cite{Mitra23}. Here, we render expectation scale invariant to system size by defining $\mathbf{X}\equiv \frac{2}{N}\langle \mathbf{J}\rangle$, and of these we consider the QPTs in terms of $Z=\frac{2}{N}\langle J_z\rangle$. In both QPTs, for small $s$ the system exhibits paramagnetic alignment of the spins with the transverse field; for large $s$, the system enters an increasingly ferromagnetic phase as each spin aligns with its neighbors along the longitudinal axis. These two phase transitions differ in their critical values of $s$, and whether their origins are kinematic or dynamic. 

\begin{figure}
\centering
\includegraphics[width=\linewidth]{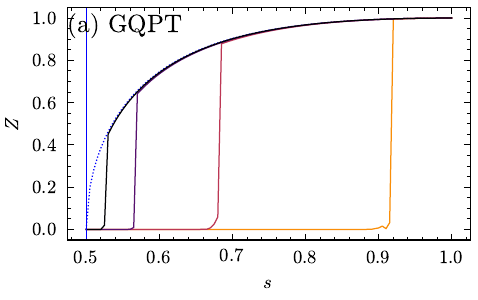}
\includegraphics[width=\linewidth]{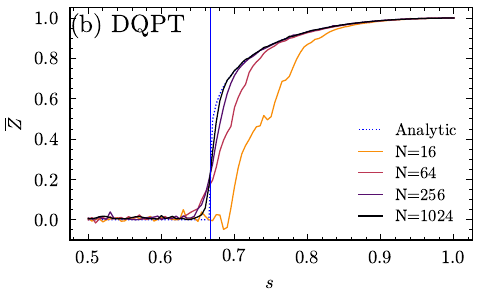}
\caption{\label{fig:QPT} The QPTs within $s\in[0.5,1]$ for $\alpha=1$ at various $N$. (a) the GQPT measured by $Z$. (b) the DQPT measured by $\overline{Z}$ with an integraion time of $T=10^5$. In both cases, as system size increases, the curves approach the analytic thermodynamic limits (blue dotted lines).}
\end{figure}

The GQPT is defined by the Hamiltonian's ground-state $\ket{n=0}$ (which generically lies on the corner of the tower of states, and thus safely within the QMBS) with critical $s_{GQPT}=\frac{1}{2}$ in the case of the LMG, shown in \cref{fig:QPT}a in blue. The GQPT's ferromagnetic phase results from two-fold degeneracy in the ground-state, corresponding to the aforementioned pitchfork bifurcation in the classical energy function $E(s;\theta,\phi)$. In the paramagnetic phase, the ground state is spin-flip symmetric, but superpositions of the degenerate ground states in the ferromagnetic phase can spontaneously break this symmetry, resulting in $\overline{\langle J_z\rangle}_{GQPT}=\bra{n=0}\hat{J}_z\ket{n=0}\neq 0$ wherein the overline signifies an infinite time-average, which is trivial for expectation values of Hamiltonian eigenstates. At finite size, the GQPT is not visible at $s=\frac{1}{2}$ itself, due to Heisenberg uncertainty and ensuing finite width of the ground-state's wavefunction covering shallow double-well potentials. Instead, the GQPT becomes measurable only at $s_{GQPT}^{(N)}>\frac{1}{2}$, where the double-well energy function becomes deep enough to distinguish the two degenerate ground-states; more accurately, $\lim_{N\to\infty}s_{GQPT}^{(N)}=\frac{1}{2}$.

Meanwhile the DQPT is defined by quench-dynamics starting in the $Z$-polarized state, $\ket{J=\frac{N}{2},M=\frac{N}{2}}$ and measuring the infinite time average $\overline{\langle J_z\rangle}_{DQPT}=\lim_{T\to\infty}\frac{1}{T}\int_0^T \langle \hat{J}_z\rangle(t)dt$. As discussed in Sec. \ref{sec:model}, the DQPT's paramagnetic phase occurs when the initial collection spin direction lies outside the separatrix, and dynamics restore the initial state's spontaneously broken spin-flip symmetry. On the other hand, in the ferromagnetic phase, the initial state is inside the separatrix and bound to one of the two wells, whereupon the spontaneously broken symmetry is not dynamically restored. The critical point is calculated by setting $E(s_{DQPT};0,0)=E(s_{DQPT};\frac{\pi}{2},0)$ and solving for $s_{DQPT}$. For the the LMG model, the critical point occurs at $s_{DQPT}=2/3$, shown in \cref{fig:QPT}b in blue. Previous work by \v{Z}unkovi\v{c} \textit{et al.}~\cite{Zunkovic18} has found that $s_{DQPT}$ remains well-defined with order-parameter $\overline{Z}$ for $\alpha\leq 2$, but using time-dependent variational principle (TDVP) evolution at smaller system sizes with coarser numerics. We will pursue larger systems and longer times with finer combs of $s$. In any case, for all finite $N$, both $s_{GQPT}$ and $s_{DQPT}$ are outside the QMBS regime determined in Sec. \ref{sec:delE}, so we cannot expect IRD to well-approximate the microstate evolution, rather, we must test the order parameter evolution separately.

The GQPT in \cref{fig:QPT}a, and DQPT in \cref{fig:QPT}b, are both faithfully represented by distilled irreps, and they approach the analytic forms in the thermodynamic limit, taken from \cite{Chinni21}. \Cref{fig:QPT} shows these QPTs for system sizes $N\in\{16,64,256,1024\}$, wherein growing system sizes approach $s_{GQPT}^{(N)}\to \frac{1}{2}$ and $s_{DQPT}^{(N)}\to \frac{2}{3}$ as expected. Moreover the derivative $\frac{d}{ds}\overline{Z}$ becomes smooth at large $N$, and sharply increases around the critical point. Although not shown in \cref{fig:QPT}, the distilled QPTs' behavior are largely constant across $\alpha$ for a given $N$, thanks to Kac normalization.

That the GQPT cleaves so tightly to its limit-behavior under the LMG is well-explained by the spectrum; the ground-state of $\hat{H}(s,\alpha)$ remains extremely close to, or exactly within, $\mathcal{H}_{N/2}$ across nearly all parameters. This is because both operator components, the transverse field and longitudinal interaction, have their expectation values extremized by symmetric states. Thus the ground-state (and, sometimes, the maximally excited state) are both well represented under distillation even when the QMBS is otherwise comprehensively broken. However the DQPT has no such spectral alibi, and so illustrates a recurring point appearing across the field of many-body physics, that certain classes of macrostatic observables remain robust even in chaotic systems that should be resistant to approximation, and even under highly reductive and seemingly unstable approximation schemes.

Both of the QPTs are explained by the classical energy $E(s;\theta,\phi)$, which lives in the phase space representation of the symmetric subspace $\mathcal{H}_{N/2}$. That the DQPT and GQPT remain largely unchanged under distillation and truncation at finite $\alpha$ suggests not only that the distilled Hilbert space abides by a kind of phase-space, but that the energy landscape of said phase-space changes only perturbatively in the long-range interaction regime, from its collective limit. But of course the distilled subspace is not closed under long-range dynamics, and so in addition to energy functions, we must consider the stability of states in phase space.

\subsection{Dynamics of Nonclassical States}\label{sec:NCS}
\begin{figure}
\centering
\includegraphics[width=\linewidth]{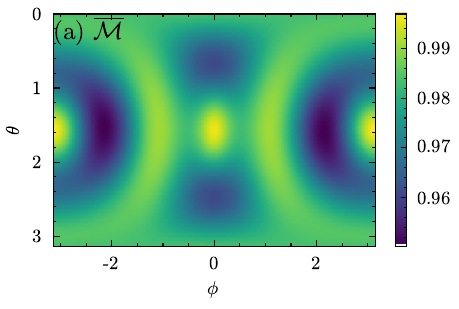}
\includegraphics[width=\linewidth]{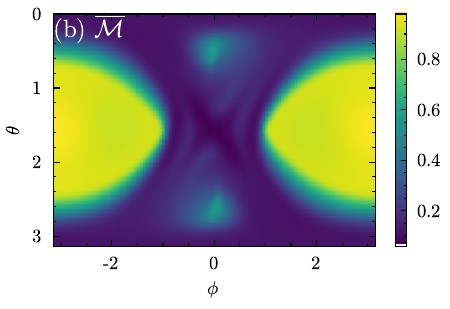}
\caption{\label{fig:SCS} Time-averaged SCS Loschmidt echoes $\overline{\mathcal{M}}(\theta,\phi)$, from \cref{eq:SCSLoschmidt}, for $N=100$, $\alpha=0.4$, and a total averaging time $T=20$ as a function of the direction of an initial spin coherent state $\ket{\theta, \phi}$. (a) Within the QMBS regime at $s=0.4$, all initial polarizations maintain high fidelity over short evolutions. (b) At the DQPT critical point $s=s_{DQPT}$, initializations near the separatrix and unstable fixed point fail to converge in approximate evolution.}
\end{figure}
\begin{figure}
\centering
\includegraphics[width=1\linewidth]{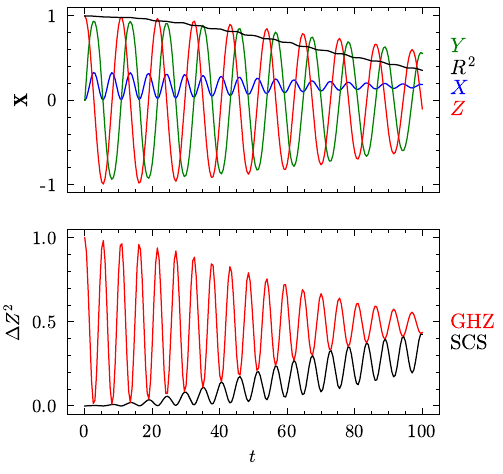}
\caption{\label{fig:Nonclassical} Dimensionless collective observable of distilled evolution, $\mathbf{X}$, at $N=400$, $(s,\alpha)=(0.4,0.8)$, initialized in $Z$-polarized states. Top: $X$, $Y$, $Z$, and $R^2\equiv||\mathbf{X}||^2$ (black). Bottom: $\Delta Z^2\equiv \frac{4}{N^2}(\langle J_z^2\rangle-\langle J_z\rangle^2)$ initialized in SCS (black), and GHZ (red).}
\end{figure}

The IRD allows for the models of dynamics beyond mean field classical approximation. Two canonical examples of nonclassical states, squeezed and cat states, are both generated by the Ising and LMG family of Hamiltonians. Because dynamical instability and chaos cause slightly perturbed points in phase space to diverge in their evolution, these regimes can efficiently split classical wavefunctions into quantum superpositions. Beyond the pitchfork bifurcation for $s>s_{GQPT}$, the unstable fixed point at $(\theta,\phi)=(\frac{\pi}{2},0)$ or at $\ket{\uparrow_x}$ propels that initialization along the separatrix, squeezing and then splitting it into a cat state~\cite{Munoz23}. Since at $s=s_{DQPT}$, said separatrix passes through the $Z$-poles, then $\hat{H}(\frac{2}{3},0)$ acting on $\ket{\theta=\frac{\pi}{2},\phi=0}$ can generate a $Z$-polarized spin-GHZ state: $\frac{1}{\sqrt{2}}(\ket{\frac{N}{2},\frac{N}{2}}+\ket{\frac{N}{2},-\frac{N}{2}})$.

But instability and chaos, which are useful for rapid state development, also combine into ergodic decay of coherences. For all $s\in(0,1)$ and all $\alpha>0$, then $\hat{H}(s,\alpha)$ is generically chaotic; but only at $s\geq \frac{1}{2}$ does it gain an unstable phase space in its symmetric manifold. To illustrate the implication of the GQPT in the breakdown of the QMBS and ensuing failure of approximation, we turn to Loschmidt analysis. For two Hamiltonians $\hat{H}\approx \hat{H}'$ and an initial state $\ket{\psi}$, the Loschmidt echo is defined,
\begin{equation}
 \mathcal{M}(\psi,t)\equiv |\bra{\psi}e^{i\hat{H}'t}e^{-i\hat{H}t}\ket{\psi}|^2.
\end{equation}
The timescale and initialization subspace over which $\mathcal{M}\approx 1$ determines the conditions under which $\hat{H}'$ can dynamically reproduce the fine-grained structure of $\hat{H}$. 

Consider $\hat{H}=\hat{H}_D(s,\alpha)$, the distilled Hamiltonian, $\hat{U}_{QMBS}(t)\equiv \sum_{n'}e^{-iE_n't}\ket{n'}\bra{n'}\approx e^{-i\hat{H}'t}$, the time-evolution map spanned by the perturbed scar states constructed in \cref{eq:pert1,eq:pert2} which approximates a unitary over that subspace, and an initial SCS, $\ket{\theta,\phi}$. We can approximately decompose $\ket{\theta,\phi}\approx\sum_{n'}\braket{n'}{\theta,\phi}\ket{n'}$, and accordingly define
\begin{align}
\mathcal{M}(\theta,\phi;t)=&|\bra{\theta,\phi}\hat{U}^\dag_{QMBS}(t)e^{-i\hat{H}_D(s,\alpha) t}\ket{\theta,\phi}|^2\nonumber\\
=&|\sum_{n'}\braket{\theta,\phi}{n'}e^{iE_n' t}\bra{n'}e^{-i\hat{H}_D t}\ket{\theta,\phi}|^2\label{eq:SCSLoschmidt}\\
\approx &|\bra{\theta,\phi}e^{i\hat{H}'t}e^{-i\hat{H}_D(s,\alpha) t}\ket{\theta,\phi}|^2.
\end{align}
Note that all the population of $\ket{\theta,\phi}$ outside the span of the perturbed scars is projected away, by $\hat{U}_{QMBS}$ that acts only on a subspace, suppressing the Loschmidt echo. As discussed in Sec. \ref{sec:delE}, for $t\ll \frac{1}{\Delta E}$, $\mathcal{M}(n',t)\approx 1-(\Delta E_{n'} t)^2$. That is, the energy variance forms the leading order term in Loschmidt echo decay. We time-average the Loschmidt echo as $\overline{\mathcal{M}}(\theta,\phi)$, and plot the results in \cref{fig:SCS}.

\Cref{fig:SCS} demonstrates the impact of the boundary shown in \cref{fig:varEParam} on dynamical fidelity across the Bloch sphere. \Cref{fig:SCS}a shows $\overline{\mathcal{M}}(\theta,\phi)$ for parameters with uniformly low $\Delta E$, indicating all trajectories in the symmetric subspace are well-approximated over considerable time. But \cref{fig:SCS}b shows that at $s=s_{DQPT}$, the unstable fixed point and the separatrix crossing over it have $\overline{\mathcal{M}}\approx 0$, indicating rapid decay. The two wells above and below the unstable fixed point, and especially the high-energy region on either side, have far stronger echoes. In fact, the high-energy region suggests a remaining scarred subspace that doesn't fully cover the symmetric subspace. This and other corroborating findings imply that the breach in the QMBS, and consequently the non-convergence of IRD's approximation along specific trajectories, results from the combination of chaos from localization and instabilities in phase space beyond the GQPT. 

The discrepancy between the critical value associated with the breakdown in QMBS ($s_c\approx 0.42$) and the point of ground-state quantum phase transition ($s_{GQPT}=0.5$) can be understood as follows. Recall that the GQPT is a phase transition in a macroscopic parameter which is insensitive to fine-grained structure, which would first be detected by spectral quantities like $\Delta E_n'$. Moreover, the phase-space of the long-range system, even in the thermodynamic limit, is higher-dimensional than a simple sphere for one degree of freedom, and unstable trajectories may emerge in the full phase space whose instability not seen in $(\theta,\phi)$.

Unstable trajectories in phase space permit rapid evolution to generate nonclassical states, but in chaotic systems they also breach the QMBS. Since IRD approximates the QMBS, it cannot accurately represent dynamics along these trajectories, although it still captures stable regions of phase space even in parametric regimes that include these unstable fixed points and separatrices. Moreover, irrep distillation can perfectly well represent nonclassical states, even if stable IRD dynamics can only generate them slowly. Consulting \cref{fig:Nonclassical}, states that reside mostly within the symmetric subspace, but that don't resemble spin-coherent states, commonly emerge from dynamics in the QMBS regime, visible in the shrinking $R^2$ net polarization. Likewise, the effects of long-range QMBS dynamics on prior-prepared nonclassical states, like the GHZ state in red, may come of interest, so the ability of IRD to predict these dynamics presents utility.

\section{Discussion}\label{sec:disc}
\subsection{Comparison with Other Methods}\label{sec:comp}
Irrep distillation is not the only approximate method for representing long-range interacting spin-ensembles in their near-collective scenarios. Three others are established in literature: time-dependent spin-wave theory~\cite{lerose2019,li2024}, matrix product states (MPS)~\cite{Schollwock11,Pappalardi18}, and discrete truncated Wigner approximation (DTWA)~\cite{schachenmayer2015,Khasseh20}. Compared with these formalisms, IRD has a clear use-case in the construction and dynamics of highly nonlocal/entangled collective states on generic lattice geometries, with guaranteed dynamical fidelity.

Time-dependent spin-wave theory is highly scalable, as it combines the Holstein-Primakoff and Gaussian approximations, but it inherits their modeling limitations. It cannot represent, let alone evolve, non-polarized or nonGaussian-like states such as spin-GHZ states, despite such states residing purely in the symmetric subspace. Irrep distillation retains the symmetric subspace as a Hilbert space, allowing it to both represent and sometimes generate weakly- and non-polarized collective states, as discussed in Sec. \ref{sec:NCS}. Recent work by Lerose \textit{et al.}~\cite{Lerose23} extends the spin-wave formalism for long-range spin-models into a ``rotor-magnon" picture, by retaining the standing-wave mode as a collective spin degree-of-freedom, and performing the Holstein-Primakoff transformation on all other modes. This affords their formalism access to generic Dicke states, albeit in an overcomplete non-orthogonal basis that demands periodic boundary conditions.

MPS methods are useful tensor-network computational tools for efficient solutions of ground states (density matrix renormalization group, or DMRG), and time-evolution for both short-range and long-range~\cite{Haegeman16} interactions (time-dependent variational principle, or TDVP). MPS evolution is most efficient when entanglements are short-range and shallow, with a built-in time limit as entanglement generically grows over dynamics.
But all MPS methods inherit the lattice-geometric limitation baked into its left/right-operator notation, so it struggles with most systems on multidimensional lattices, where entanglement is too thoroughgoing to truncate. Irrep distillation is agnostic to lattice geometry, whether a chain, a square grid, a cubic grid, triangular couplings, etc. The lattice geometry and dimension is encoded in the relative strengths of couplings between sites, so in two and higher dimensions, our method generically outperforms MPS methods.

DTWA represents an initial state exactly as a discrete Wigner function, then evolves it according to the Poisson bracket (rather than the Moyal bracket), as though under a classical Hamiltonian. While DTWA is as efficient as the construction of its initialization, and seems to reproduce correct order parameters in a variety of systems~\cite{Khasseh20}, it is an uncontrolled approximation. DTWA isn't guaranteed to converge and only justified heuristically post-facto, and so requires independent verification. Irrep distillation's grounding in perturbation theory and its connection to QMBS provides it with guarantees and internal checks via self-consistency metrics (e.g., energy variances $\Delta E$ and Loschmidt echoes $\mathcal{M}$).

\subsection{Two-Body Model}\label{sec:2BM}
IRD provides a new way of understanding the dynamics of long-range interacting systems in terms of minimal decomposition into subsystems and degrees of freedom. Recall that the infinite-range Hamiltonian is exactly represented by a one-body collective spin model with one degree of freedom, while the finite-range system is properly many bodies. Here we show that IRD allows us to interpret the dynamics as a few-body model, specifically with two degrees of freedom to first order in perturbation theory. Furthermore this interpretation reintroduces a new tensor-product decomposition of degrees of freedom, with the mathematical benefit of well-defined entanglement. A standard quantification of entanglement in many-body systems is the entanglement entropy from a half-chain partition, but whenever this entropy saturates due to volume-law entanglement, it becomes intractable to calculate at large $N$. The entropy following from our tensor-product decomposition saturates according to an area-law, and we show that the low entanglement generated dynamically in the few-body picture arises from the optimality of the decomposition.

\begin{figure}
\centering
\includegraphics[width=1\linewidth]{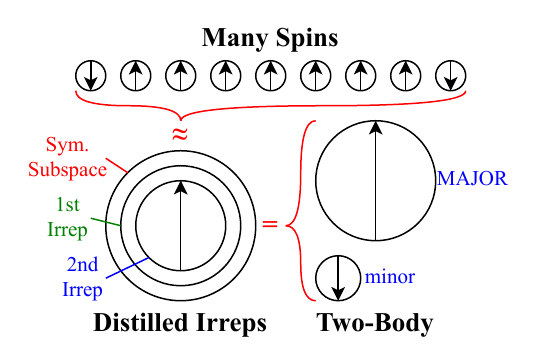}
\caption{\label{fig:TBMap}An ensemble of identical half-spin particles, represented in the collective-spin basis, may have its state truncated into a nondegenerate set of distilled irreps. These distilled irreps then form the exact collective-spin basis of a pair of particles, which each encompass nonlocal degrees of freedom from the many-body ensemble.}
\end{figure}

The distilled, truncated Hilbert space of irreps $\mathcal{H}_D\equiv\bigoplus_{\delta=0}^2 \mathcal{H}_{N/2-\delta}$ defined in \cref{eq:calHD} has dimension $||\mathcal{H}_{D}||=3(N-1)$, and by inspection, maps exactly onto a two-body Hilbert space $\mathcal{H}_D=\mathcal{H}_1\otimes\mathcal{H}_{N/2-1}$ via the Clebsch-Gordan series, analogous to \cref{eq:CGSeries1,eq:CGSeries2} and shown diagrammatically in \cref{fig:TBMap}. That is, for all $\ket{m,M}\equiv \ket{1,m}\otimes\ket{\frac{N}{2}-1,M}$ such that $\ket{1,m}\in\mathcal{H}_1$ and $\ket{\frac{N}{2}-1,M}\in\mathcal{H}_{N/2-1}$, then
\begin{align}
\ket{\frac{N}{2}-\delta,M}&=\sum_{m=-1}^1 C^{N/2-\delta,M}_{N/2-1,M-m;1,m}\ket{m,M-m},\\
\ket{m,M}&=\sum_{\delta=0}^{2}C^{N/2-\delta,M+m}_{N/2-1,M;1,m}\ket{\frac{N}{2}-\delta,M+m},
\end{align}
which appropriately has Hilbert dimensions $||\mathcal{H}_1||=3$ and $||\mathcal{H}_{N/2-1}||=N-1$. As \cref{fig:TBMap} shows, through this mapping from a trio of subspaces onto a pair of subsystems (the ``major" spin $\ket{\frac{N}{2}-1,M}$ and ``minor" spin $\ket{1,m}$), we can interpret the effect of IRD as partitioning the $N$ spin-half particles into a set of $N-2$ and a set of 2. This partition groups together the sets of maximum co\"operation so that the $N-2$ spin-halves are approximately represented by a single spin-$\frac{N}{2}-1$ major particle, while the remaining 2 spin-halves are approximated by a spin-1 minor particle.

The major and minor particles are nonlocal degrees of freedom, in contrast to the local spin-degrees of freedom in the computational basis. Whereas each subsystem $\mathcal{H}_{1/2}^{(j)}\subset \mathcal{H}_{1/2}^{\otimes N}$ encodes the full information accessible by measurements on the spin at the $j^{\text{th}}$ site and only that spin, the minor particle corresponds superpositionally to the information in all spins. The major and minor particles are nonlocal degrees of freedom, in the same way that wave-number degrees of freedom under Fourier transforms, or parity/phase degrees of freedom in the Bell basis. 

By the same token, the lowest order of permutation-asymmetry and non-collective behavior is reflected in asymmetries between the major and minor particle states, visible in their entanglement. Considering the two-body basis and tracing over either the major or minor particle, $\hat{\rho}_M=\Tr_m(\hat{\rho})$, $\hat{\rho}_m=\Tr_M(\hat{\rho})$ we achieve an alternative (generically lesser) entanglement entropy,
\begin{equation}
 S_{TB}=-\Tr(\hat{\rho}_M \ln\hat{\rho}_M)=-\Tr(\hat{\rho}_m \ln\hat{\rho}_m).
\end{equation}
We compare this to the entanglement entropies calculated by tracing over any two {\em local} degrees of freedom in the exact system. For spin-sites $j,k\in[1,N]$ and density matrix $\hat{\rho}:\mathcal{H}_{1/2}^{\otimes N}\to \mathcal{H}_{1/2}^{\otimes N}$,
\begin{equation}
 \hat{\rho}_{\neg j,k}\equiv\Tr_{j,k}(\hat{\rho}),\quad S_{\neg j,k}=-\Tr(\hat{\rho}_{\neg j,k}\ln\hat{\rho}_{\neg j,k}),
\end{equation}
in which there are $\binom{N}{2}$ unique values of $(j,k)$. Thus we define maximum, mean, and minimum entropies across the ensemble of computational-basis partial traces,
\begin{align}
 S_{max}&=\max_{j,k}S_{\neg j,k},\\
 \overline{S}&=\frac{2}{N(N-1)}\sum_{j,k}S_{\neg j,k},\\
 S_{min}&=\min_{j,k}S_{\neg j,k},
\end{align}
and measure these entropies and $S_{TB}$ for the QMBS of the distilled Hamiltonian.
\begin{figure}
\centering
\includegraphics[width=1\linewidth]{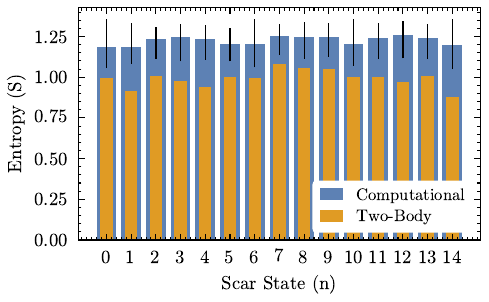}
\caption{\label{fig:EntropyTB}Entanglement entropies from partial traces over $N-2:2$ partitions, for each eigenstate $\ket{n}$ in the QMBS of a 14-spin system. Computational entropies show the range and mean of $S_{j,k}$ (blue with bounds). Two-body entropy $S_{TB}$ shown inside, always less than every computational entropy for the same state.}
\end{figure}

\Cref{fig:EntropyTB} shows this comparison of entropies, demonstrating that across the QMBS, $S_{TB}\leq S_{min}$, suggesting that the two-body mapping effectively selects the partition $N-2:2$ across which there is the {\em least} entanglement, optimizing over both local and nonlocal degrees of freedom. The optimization over nonlocal degrees of freedom is paramount; were the two-body model constrained to assign two specific sites $(j,k)$ to fall under the purview of $\mathcal{H}_1$, then the two-body entropy would be bound below, $S_{TB}=S_{min}$. By the monogamy of entanglement, consequently the minor and major particles each enclose the most collectively co\"operating (ergo entangled) degrees of freedom.
 
Discrepancies between $S_{TB}$ and $S_{min}$ arise from two sources, primarily that the two-body mapping isolates and localizes correlations which are scrambled in the computational basis, thereby presenting an infimum (but not necessarily minimum) for entanglement from all possible $N-2:2$ partitions. Because partial traces over the computational basis only respect local degrees of freedom, if the greatest correlations are nonlocal, then no choice of $(j,k)$ minimizes the entropy. The two-body model is constructed from the first-order expansion of the symmetric subspace $\mathcal{H}_{N/2}$ under the Hamiltonian $\hat{H}(s,\alpha)$. This means the first coherences in symmetric initial states to break down dynamically are those between the major and minor particle, and so this subsystem decomposition generates the least entanglement under $\hat{H}(s,\alpha)$ out of any basis. In the limit case of the LMG at $\alpha=0$, the system restores permutation symmetry, and $(S_{max}=\overline{S}=S_{min}=S_{TB})|_{\alpha=0}$, as is expected when all $N-2:2$ partitions, whether local or nonlocal, are equivalent. Secondarily, disagreement may be systematic between the distilled $\hat{H}_D$ and the exact $\hat{H}$, as when outside the QMBS regime. In this case, $S_{TB}\not \leq S_{min}$ because perturbation theory is nonconvergent and the distillation-approximate QMBS is a poor approximation.

\section{Conclusion and Outlook}\label{sec:C&O}
In this investigation we have defined a new approach to study quantum many-body systems with long-range interactions that truncates the exponentially large Hilbert space in a manner that efficiently captures the desired dynamics. Our scheme is motivated by quantum many-body scars (QMBS) which yield ordered dynamics on non-integrable many-body spin systems analogous to their integrable collective counterparts, and we isolate the QMBS by a special truncation of Hilbert space. This truncation, via ``irreducible representation distillation" (IRD) exploits irrep degeneracy to solve for ``bright" irreps that dress the collective system, thereby reproducing the first-order perturbation theory of the QMBS with minimal matrix elements. We deployed IRD on the long-range interacting transverse field Ising Hamiltonian, since its infinite-range limit is the integrable LMG. Using Loschmidt echoes on the perturbation-theoretic QMBS, we tested the self-consistency of IRD, finding a clear set of parametric bounds wherein the perturbed scars are all near-eigenstates under distilled dynamics. We related these bounds to the existence and collapse of the QMBS itself, as a product of unstable trajectories in the classical phase space resulting from the GQPT. Even outside these parametric bounds, we found that distilled dynamics reproduce the QPTs of the LMG, corroborating prior results~\cite{Zunkovic18}. 

The bounds on the applicability of IRD is closely related to the existence of unstable fixed points in the classical limit, and the range of interactions. We find that critical coupling strength $s\leq s_c^{(N)}$ for $\lim_{N\to\infty}s_c^{(N)}=0.42$ (approaching the limit from above) and $\alpha\leq 1$, agree with the bounds $\alpha\leq d$ given by~\cite{Lerose23} for lattice dimension $d$. In addition, the form of the breakdown in QMBS at $s>s_c$ fulfills that paper's prediction, that approximate energy-degeneracies between states of neighboring spin-number levels would lead to hybridization in the corresponding eigenstates, and consequent breakdown of the level separation. The correspondence of this QMBS collapse with the GQPT critical point extends to the individual hybridizing eigenstates, being those with population along the phase-space separatrix. The position of the energy-region wherein this degeneracy arises varies smoothly with $s\in [0.5,1]$ at $0<\alpha\ll 1$, vindicating the prior spectral description of such breakdowns occurring because of resonances in two smoothly varying characteristic frequencies of the Hamiltonian. However, ~\cite{Lerose23} claims that these resonances occur only at exceptional parameter values, whereas we found them endemic to half the considered parameter space. Moreover, they argued that systematic collapse of QMBS chiefly results from classical chaos according to Kolmogorov-Arnold-Moser theory, but we have found that mere exceptional unstable trajectories, rather than a fully unstable phase-spatial bulk, breaches the QMBS.

The method of IRD, and its interpretation as a reduction to a few spin degrees-of-freedom, is agnostic of lattice-geometry, and remains mathematically well-defined for any Hamiltonian with a finite number of permutation symmetry-breaking terms, regardless of the support of those operators or the forms of accompanying symmetric terms. If couplings between the symmetric subspace and lower subspaces can be attributed to a finite number of low-weight operators modulated by localizing functions, then IRD can optimize a set of nondegenerate or low-fold degenerate irreps which exclusively couple via said localizing functions. Furthermore, IRD is a controlled approximation thanks to its correspondence with perturbation theory, and for any finite $N$, a sufficiently high order of IRD (described below) will converge to the full Krylov subspace of the QMBS. Its region of applicability is efficiently testable. Finally, IRD can represent and even slowly generate nonpolarized, nonclassical, non-Gaussian states because it retains the mathematical structure of a (truncated) Hilbert space. This combination of features makes IRD somewhat unique among the other approximations in Sec. \ref{sec:comp}, and these differences have profound implications for simulation.

To illustrate this, consider the application of IRD to control the many-body state in the symmetric subspace, but with finite range interaction. Such control has been of great interest recently for applications to metrology, with the production of spin squeezed states in arrays of Rydberg atoms in optical tweezers~\cite{eckner2023,bornet2023}. A $2\sim3$D lattice of atoms described by a spin XX model will exhibit QMBS with a strong overlap with the symmetric subspace and large energy gap. Through application of a time-dependent driving field, the system is in principle ``controllable," meaning that one can implement any unitary map within the symmetric subspace. To achieve this, one can employ quantum optimal control, but this requires the ability to efficiently integrate the time-dependent Schr\"{o}dinger equation. IRD provides the formalism necessary to do so. MPS would struggle to solve the dynamics of even modest 2D lattices. Time-dependent spin-waves could not represent non-Gaussian target states. DTWA would not permit fine-grained analysis of which subspaces afford high-fidelity simulation under highly nonclassical dynamics. We will explore this application in future work.

The IRD method presented here corresponds to only a first-order perturbation theory; one can conceive of higher-orders of IRD which acquire the minimal matrix elements for corresponding orders of perturbation theory. Similarly, permutation-symmetric operators are generically difficult experimentally, so Hamiltonians which include multiple weakly permutation-asymmetric components merit their own treatment under distillation. These extensions, along with proofs of their hierarchy of couplings and generalized few-body interpretations is the subject of forthcoming research.

Finally, the first-order approximation of a many-body Hamiltonian from a one-body ansatz is a few-body model (in this case, two-body) inviting an interpretation of dynamics through a semi-collective lens, such as the extension of QPTs directly from the few-body phase-space. The LMG's quantum phase transitions emerge as a consequence of bifurcations and separatrices in its classical energy function. These critical points are also visible in the quantum phase-spatial analogue, the Weyl symbol $W_{LMG}(\theta,\phi)$. Extensions of Weyl symbols to multiple irreps are provided by Klimov and Romero~\cite{Klimov08,Klimov17}, but cannot accommodate irrep degeneracy. By distilling and truncating, we can derive approximate Weyl symbols for the long-range Hamiltonian, both in the irrep basis and by mapping onto a doubled phase-space $(\Theta,\Phi;\theta,\phi)$ corresponding to the two-body model. These Weyl symbols may reproduce the critical behavior which explains the preservation of quantum phase transitions into the long-range regime, or may reveal new phase transitions which are irreducibly multi-body.
\section{Acknowledgements}
We thank Anupam Mitra for helpful discussions. We acknowledge funding from the US National Science Foundation, with grant PHY-2210013.
\bibliography{Bibliography}
\appendix
\section{Derivation of Hamiltonian Tensor Weights}\label{app:Fsol}
Here, we provide a more structured method for analytically solving $F_0,F_1,F_2$ as they appear in \cref{eq:Ham2}. First and easiest, we see by examination that \cref{eq:F1} simplifies to
\begin{equation}
 F_1[J]=\Tr(\hat{J}_x \hat{T}^{(1)}_1[J,u;J,u]),
\end{equation}
and further simplification comes from constraining \cref{eq:sphT} to demand $(J,u)=(J',v)$, for which the low-ordered Clebsch-Gordan coefficient provides a simple closed-form expression: $C_{J,M;1,1}^{J,M+1}=\sqrt{\frac{(J-M)(J+M+1)}{2J(J+1)}}$, which is clearly proportional to the matrix elements of $\hat{J}_+$. A mechanical calculation along these lines quickly reveals that
\begin{equation}
 F_1[J]=\sqrt{\frac{J(J+1)(2J+1)}{6}}.
\end{equation}
Such closed forms will prove unavailable for the general cases $F_0$ and $F_2$, however.

Canonically, the traces in \cref{eq:F0,eq:F2} are solved via the Clebsch-Gordan series of angular momentum addition, outlined in \cref{eq:CGSeries1,eq:CGSeries2}. This is intractable at larger system sizes, representing a bottom-up approach to acquire collective states from smaller spin ensembles. If we concern ourselves only with the symmetric subspace and other large irreps, we need a top-down analytic expression for $F_0$ and $F_2$. Towards this end, we solve
\begin{align}
 \Tr(\hat{\sigma}_z^{(j)}\hat{\sigma}_z^{(k)}\hat{T}^{(0)}_0[J,u;J,v])=\nonumber\\
 \sum_M \bra{J,M}\hat{\sigma}_z^{(j)}\hat{\sigma}_z^{(k)}\ket{J,M}\bra{J,M}\hat{T}^{(0)}_0\ket{J,M},\label{eq:Tr0}\\
 \Tr(\hat{\sigma}_z^{(j)}\hat{\sigma}_z^{(k)}\hat{T}^{(2)}_0[J,u;J',v])=\nonumber\\
 \sum_M \bra{J',M}\hat{\sigma}_z^{(j)}\hat{\sigma}_z^{(k)}\ket{J,M}\bra{J,M}\hat{T}^{(2)}_0\ket{J',M},\label{eq:Tr2}
\end{align}
wherein the matrix elements of $\hat{T}^{(k)}_q$ can be specified by placing constraints on \cref{eq:sphT}, while those of $\hat{\sigma}_z\otimes\hat{\sigma}_z$ we proceed to solve here.

The aforementioned matrix elements in Dicke space are given for $J'\leq J\leq J'+2$ as $\bra{J,M}\hat{\sigma}_z^{(j)}\hat{\sigma}_z^{(k)}\ket{J',M}=$
\begin{equation}
\frac{\bra{J,J'}\hat{J}_+^{J'-M}\hat{\sigma}_z^{(j)}\hat{\sigma}_z^{(k)}\hat{J}_-^{J'-M}\ket{J',J'}}{||\bra{J,J'}\hat{J}_+^{J'-M}||\cdot||\hat{J}_-^{J'-M}\ket{J',J'}||},
\end{equation}
since up to vector norm, general Dicke states are equivalent to the stretch states of their irreps, under the action of lowering operators. We simplify the above by taking a normal-order operator form so the ladder operators act on the bra, and the Paulis act on the stretch-state ket. We use $SU(2)$ commutation relations to acquire this normal ordering, and the recursion relation of ladder operators to renormalize the Dicke states,
\begin{align}
 \hat{J}_+^{n}\hat{\sigma}_z^{(j)}\hat{\sigma}_z^{(k)}\hat{J}_-^{n}&=
 \hat{J}_+^{n}\hat{J}_-^{n}\hat{\sigma}_z^{(j)}\hat{\sigma}_z^{(k)}\nonumber\\
 &-2n\hat{J}_+^{n}\hat{J}_-^{n-1}(\hat{\sigma}_z^{(j)}\hat{\sigma}_-^{(k)}+\hat{\sigma}_-^{(j)}\hat{\sigma}_z^{(k)})\nonumber\\
 &+4n(n-1)\hat{J}_+^{n}\hat{J}_-^{n-2}\hat{\sigma}_-^{(j)}\hat{\sigma}_-^{(k)},\label{eq:normalorder}\\
 ||\hat{J}_-^{n}\ket{J,M}||&=\sqrt{\frac{(J-M+n)!(J+M)!}{(J-M)!(J+M-n)!}}.\label{eq:laddernorm}
\end{align}
Application of these identities yields the general expression for Dicke-space matrix elements of dual Pauli matrices,
\begin{widetext}
\begin{align}
 \bra{J,M}\hat{\sigma}_z^{(j)}\hat{\sigma}_z^{(k)}\ket{J',M}=
 \sqrt{\frac{(J-M)!(J'+M)!(J+J')!}{(J'-M)!(J+M)!(2J')!(J-J')!}}\left(\sum_{\Vec{s}}(-1)^{s_j+s_k}\braket{J,J'}{\Vec{s}}\braket{\Vec{s}}{J',J'}\right.\nonumber\\
 \left.-\frac{2(J'-M)}{\sqrt{(J+J')(J-J'+1)}}\sum_{\Vec{s}}\braket{J,J'}{\Vec{s}}\bigl((-1)^{s_j}s_k\braket{\Vec{s}\oplus 1_k}{J',J'}+(-1)^{s_k}s_j\braket{\Vec{s}\oplus 1_j}{J',J'}\bigr)\right.\nonumber\\
 \left.+\frac{4(J'-M)(J'-M-1)}{\sqrt{(J+J')(J+J'-1)(J-J'+2)(J-J'+1)}}\sum_{\Vec{s}}s_js_k\braket{J,J'}{\Vec{s}}\braket{\Vec{s}\oplus (1_j, 1_k)}{J',J'}\right)\label{eq:MatElem}
\end{align}
\end{widetext}` 
for computational basis states $\ket{\Vec{s}}$ for $\Vec{s}\in\mathbb{B}^{\otimes N}\cong \{0,1\}^{\otimes N}\cong\{\uparrow,\downarrow\}^{\otimes N}$. In \cref{eq:MatElem}, we have an expression for Pauli operators in the irrep basis which is nearly closed-form, but for the ambiguity of the inner products between Dicke and computational states, $\braket{J,J'}{\Vec{s}}$. By closing this final ambiguity, it will become possible to analytically express Dicke-basis matrix elements of the Hamiltonian within the largest irreps, without recourse to lower irreps or recursion relations.

Consider for example $\mathcal{H}_{N/2}$, $\mathcal{H}_{N/2-1,u}$, and $\mathcal{H}_{N/2-2,u}$, since only those subspaces are directly coupled to the symmetric subspace at lowest order. In these cases, we can relate $\braket{J,J'}{\Vec{s}}$ to computational amplitudes defined in \cref{eq:compamp1,eq:compamp2,eq:compamp3}, as such:
\begin{align}
 \braket{\frac{N}{2},\frac{N}{2}}{\uparrow}^{\otimes N}&=1,\\
 \braket{\frac{N}{2},\frac{N}{2}-1}{\downarrow_j}&=\frac{1}{\sqrt{N}},\\
 \braket{\frac{N}{2},\frac{N}{2}-2}{\downarrow_j\downarrow_k}&=\sqrt{\frac{2}{N(N-1)}},\\
 \braket{\frac{N}{2}-1,\frac{N}{2}-1,u}{\downarrow_j}&=c^{(1,u)}_j,\\
 \braket{\frac{N}{2}-1,\frac{N}{2}-2,u}{\downarrow_j\downarrow_k}&=\frac{c^{(1,u)}_j+c^{(1,u)}_k}{\sqrt{N-2}},\\
 \braket{\frac{N}{2}-2,\frac{N}{2}-2,u}{\downarrow_j\downarrow_k}&=c^{(2,u)}_{j,k},
\end{align}
and so forth.

Substituting appropriate computational amplitudes for the inner products $\braket{J,J'}{\Vec{s}}$ in \cref{eq:MatElem}, then performing the mutual traces in \cref{eq:Tr0,eq:Tr2} and summing over the localizing function as in \cref{eq:F0,eq:F2}, gives closed-form expressions for the unknown matrix weights, within the concerned subspaces:
\begin{widetext}
\begin{align}
 F_0[\alpha,\frac{N}{2},0,0]&=\frac{\sqrt{N+1}}{12}(1-N),&\\
 F_0[\alpha,\frac{N}{2}-1,u,v]&=\frac{\sqrt{N-1}}{12}\left((1-N)\delta_{uv}\right.&\left.+\sum_{j,k}\frac{2}{\mathcal{N}|j-k|^{\alpha}}(c^{(1,u)}_j-c^{(1,u)}_k)^*(c^{(1,v)}_j-c^{(1,v)}_k)\right),\\
 F_0[\alpha,\frac{N}{2}-2,u,v]&=\frac{\sqrt{N-3}}{12}\Bigg((1-N)\delta_{uv}&\left.-\sum_{j,k}\frac{2}{\mathcal{N}|j-k|^\alpha}\left(\sum_m(c^{(2,u)}_{jm}+c^{(2,u)}_{km})^*(c^{(2,v)}_{jm}+c^{(2,v)}_{km})\right.\right.\nonumber\\
 &&\left.\left.+2c^{(2,u)*}_{jk}c^{(2,u)}_{jk}-2\sum_m(c^{(2,u)*}_{jm}c^{(2,v)}_{jm}+c^{(2,u)*}_{km}c^{(2,v)}_{km})\right)\right),
\end{align}
\begin{align}
 &F_2[\alpha,\frac{N}{2},0;\frac{N}{2},0]=\frac{-1}{6}\sqrt{\frac{(N+3)(N+2)(N+1)(N-1)}{5N}},\\
 &F_2[\alpha,\frac{N}{2},0;\frac{N}{2}-1,u]=\frac{-1}{2\mathcal{N}}\sqrt{\frac{(N+2)(N+1)}{30(N-1)}}\sum_{j,k}\frac{c^{(1,u)}_j+c^{(1,u)}_k}{|j-k|^{\alpha}}\label{app:F2co1},\\
 &F_2[\alpha,\frac{N}{2},0;\frac{N}{2}-2,u]=\frac{-1}{\mathcal{N}}\sqrt{\frac{N+1}{30}}\sum_{j,k}\frac{c^{(2,u)}_{jk}}{|j-k|^{\alpha}}\label{app:F2co2},\\
 &F_2[\alpha,\frac{N}{2}-1,u;\frac{N}{2}-1,v]=\frac{1}{6}\sqrt{\frac{(N+1)N(N-1)}{5(N-2)(N-3)}}\nonumber\\
 &\times\left((1-N)\delta_{uv}+\sum_{j,k}\frac{1}{\mathcal{N}|j-k|^\alpha}\left(2(c^{(1,u)*}_jc^{(1,v)}_j+c^{(1,u)*}_kc^{(1,v)}_k)+c^{(1,u)*}_jc^{(1,v)}_k+c^{(1,u)*}_kc^{(1,v)}_j\right)\right),\\
 &F_2[\alpha,\frac{N}{2}-1,u;\frac{N}{2}-2,v]=\frac{1}{2\mathcal{N}}\sqrt{\frac{N(N-1)}{30(N-4)}}\nonumber\\
 &\times\sum_{j,k}\frac{1}{|j-k|^\alpha}\left(2(c^{(1,u)*}_j+c^{(1,u)*}_k)c^{(2,v)}_{jk}-\sum_m c^{(1,u)*}_m(c^{(2,v)}_{jm}+c^{(2,v)}_{km})\right),\\
 &F_2[\alpha,\frac{N}{2}-2,u;\frac{N}{2}-2,v]=\frac{1}{6}\sqrt{\frac{(N-1)(N-2)(N-3)}{5(N-4)(N-5)}}\\
 &\times\left((1-N)\delta_{uv}-\sum_{j,k}\frac{1}{\mathcal{N}|j-k|^\alpha}\left(4c^{(2,u)*}_{jk}c^{(2,v)}_{jk}-\sum_m\left((c^{(2,u)}_{jm}+c^{(2,u)}_{km})^*(c^{(2,v)}_{jm}+c^{(2,v)}_{km})+(c^{(2,u)*}_{jm}c^{(2,v)}_{jm}+c^{(2,u)*}_{km}c^{(2,v)}_{km})\right)\right)\right).\nonumber
\end{align}
\end{widetext}
Of these, \cref{app:F2co1,app:F2co2} are reproduced in the main text as \cref{eq:F2co1,eq:F2co2}, motivating the move to irreducible representation distillation.

\section{Finite-Size Scaling Analysis}\label{app:FSSA}
Coarse-grained measurements of $\Delta E_{max}$ according to \cref{eq:energy_variance} and \cref{fig:varEParam} find a sharp increase in the worst-case approximation of Hamiltonian eigenstates by perturbation theory, at a specific point in $s$. Here we determine the form of this increase in $\Delta E_{max}$ with $s$ via fine-grained numerics around the apparent critical point, and use manual finite-size scaling analysis (FSSA) to approximate the this critical point and associated critical exponents. The result is an apparent phase-transition in the convergence of perturbation theory with $s$ in association with (but distinct from) the GQPT described in the main text.

\begin{figure}
\centering
\includegraphics[width=1\linewidth]{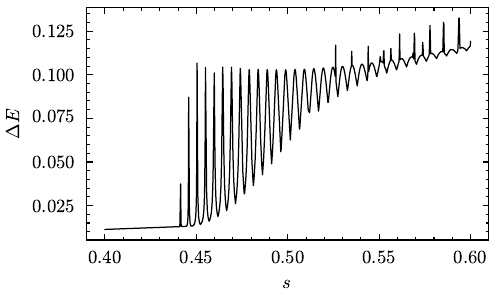}
\caption{\label{fig:varE1} The maximum energy variance of distilled eigenstates, $\Delta E_{max}$, for $N=128$ and $\alpha=0.5$, plotted over $s\in[0.4,0.6]$. At low $s$, a stable region shows uniformly low $\Delta E$, followed by a series of oscillations with sharp peaks and smoothly increasing troughs, beginning well below the apparent border of $s=0.5$ depicted in the coarse-grained \cref{fig:varEParam}b.}
\end{figure}
\Cref{fig:varE1} shows a plot of $\Delta E_{max}$ against a small parameter range of $s\in[0.4,0.6]$, corresponding to a cross-section of \cref{fig:varEParam} at $\alpha=0.5$ and $N=128$, with greater data resolution. At this scale, it becomes clear that the sharp jump in $\Delta E_{max}$ at $s\approx 0.5$ shown in \cref{fig:varEParam}b is not the only non-smooth point: in fact, $\Delta E_{max}$ depends sensitively on small changes in $s$ to determine the relative frequencies in $\hat{H}(s,\alpha)$ and therefore the quality of perturbation theory. This means that within a small neighborhood of $s\in[s'-\delta s,s'+\delta s]$, there are exceptionally lucky and unlucky values of $s$ which quicken or delay the convergence of perturbative expansions to the Hamiltonian eigensystem. However, we also see from \cref{fig:varE1} that best-case values of $\Delta E_{max}$ (i.e., local minima) seem to grow smoothly as $s$ increases across the displayed range, implying that all perturbation theories within the neighborhood $s'\pm \delta s$ will uniformly worsen as the center of that neighborhood, $s'$, crosses some critical value $s_c$.

\begin{figure}
\centering
\includegraphics[width=1\linewidth]{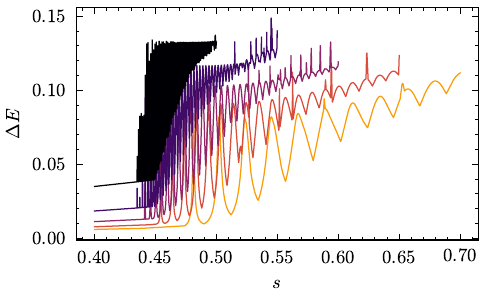}
\includegraphics[width=1\linewidth]{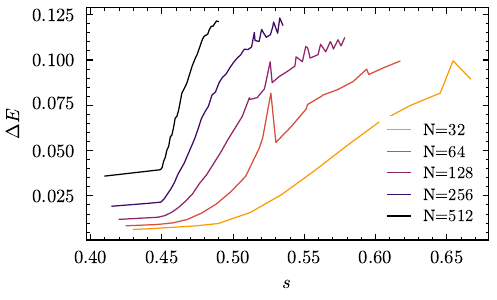}
\caption{\label{fig:varE2} Maximum energy variance $\Delta E_{max}$, for $\alpha=0.5$ and $N\in\{32,64,128,256,512\}$, plotted over small ranges in $s$. Top: full scans in $s$, showing sharp oscillations. Bottom: local minima selected from above, showing mostly smooth increase of $\Delta E$.}
\end{figure}
\begin{figure}
\centering
\includegraphics[width=1\linewidth]{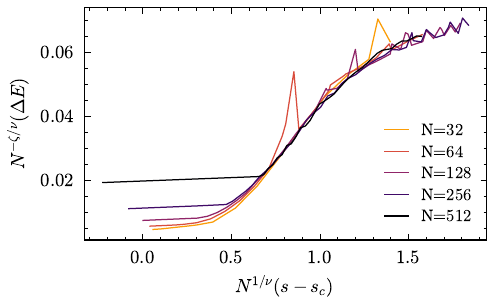}
\caption{\label{fig:varEQPT} The maximum energy uncertainty of distilled eigenstates, $\Delta E_{max}$, for $\alpha=0.5$, plotted across $s$. Local minima in $s$ show smooth increase of $\Delta E_{max}$. Manually scaled $N^{-\zeta/\nu}\Delta E_{max}$ against $N^{1/\nu}(s-s_c)$ finding $s_c\approx 0.42$, $\nu\approx 2$, $\zeta\approx 0.2$. At all observed system sizes, QMBS breakdown occurs beyond $s>0.44$, after the critical point.}
\end{figure}
The apparent point of transition from low to high $\Delta E_{max}$ depends on $N$ however, as shown in \cref{fig:varE2}, likewise the oscillations become sharper as system size increases. \Cref{fig:varE1}a finds that the bounds on low $\Delta E_{max}$ shift to lower values of $s$ at large system size, indicating that the QMBS regime becomes stricter in the thermodynamic limit. As the oscillations are not a finite-size effect, but endemic to the relationship between $\Delta E_{max}$ and $s$ at all scales, it's necessary to find a smooth subset of the data to use for FSSA, shown in \cref{fig:varE2}b which are the local minima selected from the full data-set in \cref{fig:varE2}a, and curated to exclude oscillatory regimes outside the transition of interest.

To find the critical point of this phase transition, we employ FSSA, which assumes there to be some smooth scale-dependent function $f^{(N)}$, such that $\Delta E_{max}=f^{(N)}(s)$. Without determining the form of that function, FSSA solves the best fit for a corresponding scale-invariant function $\tilde{f}$, such that $\Delta E_{max}=N^{\zeta/\nu}\tilde{f}(N^{1/\nu}(s-s_c))$, by optimizing over the critical point $s_c$ and critical exponents $\zeta,\nu$.

Because the smooth data is taken from the local minima of a non-smooth plot, it is inhomogeneous. The values of $s$ included in the dataset for one value of $N$ are not the same as those of every other. For this reason, we approximate $(s_c,\zeta,\nu)$ manually, finding the best fit in \cref{fig:varEQPT} with $s_c\approx 0.42$, $\zeta\approx 0.2$, and $\nu\approx 2$. However, note that this critical point lies well below the first oscillations in $\Delta E_{max}$ at any system size. Solving the critical point and exponents with greater precision, and determining the relationship between the critical point and actual breakdown of QMBS, remains for future research.
\end{document}